\documentclass[%
reprint,
showpacs,
 amsmath,amssymb,
 aps,
pra,
]{revtex4-1}

\usepackage{graphicx}
\usepackage{bm}
\usepackage{amsmath} 

\begin{document}



\title{Elliptic-type soliton combs in optical ring microresonators}

\author{Rodrigues D. Dikand\'e Bitha} \email{rodrigue.donald@ubuea.cm}
\author{Alain M.~Dikand\'e} \email{dikande.alain@ubuea.cm}
\affiliation{Laboratory of Research on Advanced Material and Nonlinear Science (LaRAMaNS)
, Department of Physics, University of Buea, P. O. Box Buea 063,Cameroon}  

%
%

\date{\today}

\begin{abstract}
Soliton crystals are periodic patterns of multi-spot optical fields formed from either time or space entanglements of equally separated identical high-intensity pulses. These specific nonlinear optical structures have gained interest in recent years with the advent and progress in nonlinear optical fibers and fiber lasers, photonic crystals, wave-guided wave systems and most recently optical ring microresonator devices. In this work an extensive analysis of characteristic features of soliton crystals is carried out, with emphasis on their one-to-one correspondance with Elliptic solitons. In this purpose we examine their formation, their stability and their dynamics in ring-shaped nonlinear optical media within the framework of the Lugiato-Lefever equation. The stability analysis deals with internal modes of the system via a $2\times2$-matrix Lam\'e type eigenvalue problem, the spectrum of which is shown to possess a rich set of boundstates consisting of stable zero-fequency modes and unstable decaying as well as growing modes. Turning towards the dynamics of Elliptic solitons in ring-shaped fiber resonators with Kerr nonlinearity, first of all we propose a collective-coordinate approach, based on a Lagrangian formalism suitable for Elliptic-soliton solutions to the nonlinear Schr\'odinger equation with an arbitrary perturbation. Next we derive time evolutions of Elliptic-soliton parameters in the specific context of ring-shaped optical fiber resonators, where the optical field evolution is tought to be governed by the Lugiato-Lefever equation. By solving numerically the collective-coordinate equations an analysis of the amplitude, the position, the phase of internal oscillations, the phase velocity and the energy is carried out and reveals a complex dynamics of the Elliptic soliton in ring-shaped optical microresonators.
\end{abstract}

\maketitle


\section{Introduction\label{sec:1}}

Since the observation of solitons in optical fibers and fiber lasers \cite{ki1}, the optical communication technology has experienced tremendous progress along with a wealth of modern techniques for information processings and data manipulations. This progress led, among others, to nowaday techniques of optimized data compression out of which the techniques of time, wavelength, polarization and phase multiplexings \cite{r6}. In fiber communication technology however, the spatial dimension has been a relatively less relevant issue until the revolution triggered by the advent of photonic crystals \cite{ki1,johan,ki1}, and the recent appearance of optical frequency combs (OFCs) in fiber lasers \cite{r7,r8}. On this last point photonic crystals can support hundreds of spatial modes, a feature which could be exploited for the propagation of OFCs in form of space-division multiplexed (SDM) signals. Indeed the development of systems using space-division multiplexing in photonic crystals, offers a promising technological tool for increasing the capacity of optical networks, namely by increasing the volume of information per fiber. Note that fibers supporting several spatial modes provide the most effective means
to increase the fiber-based information capacity, given that the capacity increases with the number of modes in the absence of mode-dependent losses.\\
OFCs \cite{r7,r8} are the most recent revolution in the fiber laser technology, these are discrete high-intensity optical spectra of equidistant phase-locked lines extending over a broad spectral range \cite{r7}. Since their discovery, they have been reported in various nonlinear optical media such as semiconductor microresonators \cite{r9} and fiber-laser cavities. They offer a wide-range applications in telecommunication for the generation of high-repetition rate picosecond pulses, namely for ultra-high capacity transmission systems based on optical division multiplexing \cite{r10,r11,r12}. OFCs today are widely used in spectroscopy \cite{r7}, Astronomy, metrology, frequency synthesis, optical clocking \cite{r13,r14} and so on.\\
The mechanism by which optical soliton frequency combs are generated, as well
as their stability, have been some of most active aspects in recent theoretical and experimental researches in nonlinear optics \cite{malm,zaj,yan1}. Mode-locking has emerged as the likely mechanism, it consists of the generation of a uniform train of optical solitons
formed from a four-wave mixing process, associated with the nonlinear interaction between light and the bulk material of a whispering gallery mode microresonator. Depending on the frequency and power of the pump laser, mode-locking can lead
to the excitation of equally-spaced optical soliton lines thus forming what is usually termed "soliton Kerr combs" \cite{kart1,kart2,kart3,r3}. \\
In view of the periodic structure of optical soliton Kerr combs generated by space-division multiplexing in the microresonator cavity, one might be tempted to think of its possible connection with the so-called soliton crystals \cite{am1,am2,am3,am4}, which also has been of considerable interest in the recent past in harmonic-mode-locked fiber lasers. The idea has indeed been recently introduced in a recent experimental study by Cole et al. \cite{r2}, who reported the observation of spontaneously and collectively ordered ensembles of co-propagating solitons, whose spatial discretizations allow their temporal separations within monolithic Kerr microresonator cavities. Relevant to stress, concerning a discrete ordering of some collection of several solitons to form a soliton-crystal structure, it has been established \cite{r15} that soliton crystals can result from a spatial or temporal entanglement of equidistant pulses. Quite remarkably, the resulting crystal-like ordered soliton field obeys the cubic nonlinear Schr\"odinger equation (NLSE), i.e. the same equation governing the spatiotemporal evolution of its elementary (i.e. pulse) constituents. On the other hand, the dynamics of soliton combs in Kerr microresonators are thought to be governed by the Lugiato-Lefever equation (LLE) \cite{r32}, which is actually a perturbed NLSE. Therefore whether soliton-crystal structures and soliton Kerr combs are connected, and if so in which way they could be connected, is a relevant issue for a better understanding of the process underlying the soliton frequency comb generation and its robustness vis-\`a-vis external factors such as cavity losses, external detunings and so on.\\
In this work we propose an extensive analysis of the above issue by addressing three key aspects, namely the build-up (i.e. formation) mechanism and characteristic features of the soliton-crystal structure, its stability vis-\`a-vis scattering with continuous-wave radiations, and finally the effects of cavity loss and a detuning pump on the soliton-crystal profile, within the framework of the LLE. \\
In Sec.~\ref{sec:2}, we present the perturbed NLSE specifically named LLE and discuss its soliton solutions in the absence of perturbations. Thus we first consider the artifical pulse-lattice pattern proposed by Herr et al. \cite{r16}, represented as a discrete series of fundamental sech-shaped pulses in the angular-coordinates representation. We show that this artificial structure is equivalent to a space-division multiplexing of identical pulses \cite{r15,r19,r17,r1}, and obeys the same NLSE governing the dynamics of its single-pulse constituents. As a sequel of the established one-to-one correspondance between the soliton-crystal structure of ref. \cite{r16} and the Elliptic-soliton solution to the NLSE, we derive characteristic parameters of the soliton crystals namely its period, amplitude and pulse separation, as a function of fundamental parameters of the individual pulse constituents. In sec. \ref{sec:3} we introduce an approach to the linear stability analysis for Elliptic solitons. The approach rests on a mapping of the soliton crystal-small amplitude waves scattering problem onto a $2\times 2$ matrix eigenvalue problem involving two coupled Lam\'e equations \cite{sack}. The matrix eigenvalue problem yields a rich and varied discrete spectrum comprising internal modes, growing modes and decaying modes. Spectral parameters of these modes, i.e. their eigenvalues and eigenfunctions, are determined analytically. In Sec.~\ref{sec:4} we discuss the dynamics of soliton frequency combs within the framework of the collective-coordinate approach for Elliptic-soliton solution to the LLE. We first construct a general collective-coordinate theory using a Lagrangian formalism for Elliptic-soliton solutions of the cubic NLSE with an arbitrary perturbation, and then derive the collective-coordinate equations for Elliptic-soliton parameters in the specific context of the LLE~. In Sec.~\ref{sec:5} these collective-coordinate equations are solved numerically with the help of a sixth-order Runge-Kutta algorithm~\cite{luth,r33}, combined with a $3/8$ Simpson rule for finite integrals, which also enables us explore some relevant characteristic features of the system dynamics namely the phase portraits and the energy of Elliptic solitons. Sec.~\ref{sec:6} is devoted to concluding remarks.

\section{\label{sec:2}Model, soliton crystals and Elliptic solitons}
\begin{figure}[h]
\includegraphics[scale=.5]{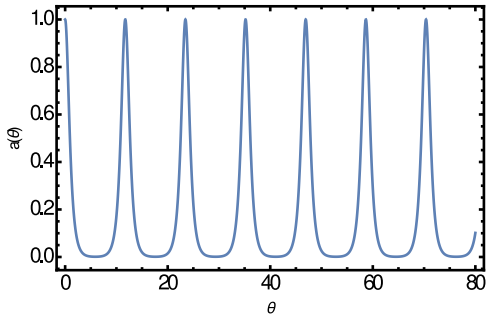}
\caption{\label{fig:f1}
(Color online) Elliptic-soliton solution to the NLSE: $\beta_2=-0.5$, $\beta=0.5$ and $\gamma=0.9$, $\kappa=0.97$.} 
\end{figure}

The dynamics of soliton Kerr combs in ring-shaped optical microresonators is usually described by the LLE, which is a perturbed NLSE given by \cite{r32}:
\begin{eqnarray}
i\frac{\partial \psi}{\partial \tau}-\frac{\beta_2}{2}\frac{\partial^2 \psi}{\partial \theta^2}+\gamma\rvert \psi
\rvert^2\psi=-i(\alpha_1 +i\alpha_2)\psi +iF,
\label{e1}
\end{eqnarray}
where $\psi=\psi(\tau,\theta)$ is the slow-varying envelope of the field, $\theta$ is the angular coordinate in the ring microresonator and $\tau$ is time. $\beta_2$ is the group-velocity
dispersion (GVD) of the microresonator, $\gamma$ is the nonlinear (i.e. Kerr) coefficient, $\alpha_1$ is the linear loss (damping term), $\alpha_2$ is the pump detuning frequency and $F$ is the pump field intensty. In all what follows we focus on the anomalous dispersion regime i.e. $\beta_2 < 0$. \\ 
In this section we are interested in the solution to the LLE for $\alpha_1=\alpha_2=F=0$. In this case Eq.~(\ref{e1}) reduces to the cubic NLSE:
\begin{eqnarray}
i\frac{\partial \psi}{\partial \tau}-\frac{\beta_2}{2}\frac{\partial^2 \psi}{\partial \theta^2}+\gamma\rvert \psi
\rvert^2\psi&=0.
\label{e2}
\end{eqnarray}
Seeking for stationary solutions to the above equation, we assume an optical field $\psi(\tau,\theta)$ of the following form \cite{r15}:
\begin{eqnarray}
\psi(\tau,\theta)=a(\theta)exp(i\beta \tau),
\label{e3}
\end{eqnarray}
where $\beta$ is the envelope modulation frequency and $a(\theta)$ is the amplitude of the field envelope assumed real. Substituting Eq.~(\ref{e3}) in Eq.~(\ref{e2}) yields:
\begin{eqnarray}
-\beta a -\frac{\beta_2}{2} \frac{\partial^2 a}{\partial \theta^2} + \gamma a^3 = 0,
\label{e4}
\end{eqnarray}
which can be transformed into the energy-integral equation: 
\begin{eqnarray}
\left(\frac{d a}{d \theta}\right)^2=-\frac{2\beta}{\beta_2}a^2+\frac{\gamma}{\beta_2}a^4+C.
\label{e5}
\end{eqnarray}
The integration constant $C$ determines profiles of the amplitude $a(\theta)$. The first physical context of interest is that of a localized profile, where the field envelope $a(\theta)$ has a vanishing shape as $\theta\rightarrow \pm\infty$ such that $C=0$. This leads to the pulse soliton solution: 
\begin{eqnarray}
a(\theta)=\sqrt{\frac{2\beta}{\gamma}}sech\left[\sqrt{\frac{-2\beta}{\beta_2}}\theta\right].
\label{e6}
\end{eqnarray}
Physically the solution (\ref{e6}) describes a high-intensity single-pulse signal of amplitude $a_0=\sqrt{\frac{2\beta}{\gamma}}$ and width $\ell_0=\sqrt{\frac{\beta_2}{-2\beta}}$. \\
When $C\neq 0$, localized structures become unstable and no pulse can form. Nevertheless the NLSE still admits nonlinear-wave solutions. Indeed, for finite nonzero values of $C$, Eq. (\ref{e5}) is an elliptic first-order differential equation admitting a solution \cite{r15,dikr,mbied1,mbied2}:
\begin{eqnarray}
a(\theta)=\frac{a_0}{\sqrt{2-k^2}}dn\left[\frac{\theta}{\sqrt{2-k^2}\ell_0},\kappa\right],
\label{e11}
\end{eqnarray}
where $dn$ is a Jacobi elliptic function of modulus $\kappa$ ($0\leq\kappa\leq 1$). The spatial profile of the envelope solution Eq. (\ref{e11}) is plotted in fig. \ref{fig:f1} for $\beta_2=-0.5$, $\beta=0.5$, $\gamma=0.9$ and $\kappa=0.97$ (values chosen for illustration).
The figure suggests a comb structure consisting of a periodic network of pulse-shaped signals, where the periodic nature of the pulse comb is actually characteristic of the Jacobi elliptic function $dn$ which is periodic in its argument with a period:
\begin{equation}
\theta_p= 2 K\sqrt{2-\kappa^2}\ell_0,
\end{equation}
with $K=K(\kappa)$ the elliptic integral of first kind. This period, which we assimilate to the pulse spatial repetition rate in the soliton-comb pattern, appears to be proportional to the width of individual pulses. It is also worth remarking that according to formula (\ref{e11}), the pulse amplitude in the pulse comb is smaller. In the limit $\kappa \rightarrow 1$ the spatial repetition rate $\theta_p$ becomes infinite and the elliptic-soliton solution (\ref{e11}) decays into the single-pulse solution (\ref{e6}). \\
To probe the inner structure of the artificial soliton-crystal pattern proposed in ref.~\cite{r16}, in the light of the two distinct exact soliton solutions to the NLSE obtained above, we remark that the authors suggested that the soliton crystal formed by SDM of pulses in microresonators could readily be represented as \cite{r16}:
\begin{eqnarray}
a(\theta)&=& \sum_{J}C_j sech\left(\sqrt{\frac{2(\omega_0-\omega_p)}{D_2}}(\theta-\theta_j)\right), \label{e7} \\
D_2&=&-\frac{c}{n_0}D_1^2 \beta_2, \label{e7aa}
\end{eqnarray}
where $\theta_j$ is the angular position of the $j^{th}$ pulse, while $\omega_0$ and $\omega_p$ are respectively the resonance frequency of the resonator and the pump Laser frequency. $n_0$ is the refractive index, $D_1$ is the free spectral range and $D_2$ is the resonator anomalous dispersion. In Eq.~(\ref{e7}) each pulse soliton is regarded as an eigenfunction with a normalized amplitude $C_j$ representing an existing mode within the soliton comb. In the specific case when pulses are identical such that $C_j=C_0$, and moreover are equally spaced such that $\theta_j=j\theta_0$, the SDM soliton crystal Eq.~(\ref{e7}) reduces to:
\begin{eqnarray}
a(\theta)=C_0 \sum_{J} sech\left(\sqrt{\frac{2(\omega_0-\omega_p)}{D_2}}(\theta-j\theta_0)\right).
\label{e8}
\end{eqnarray}
where $\theta_0$ is the pulse repetition rate. If the series (\ref{e8}) corresponds to a comb of sech-type pulses in which constituents are solutions to the NLSE, then in terms of the single-pulse solution (\ref{e6}) we can set: 
\begin{equation}
\beta=\frac{n_0}{c D_1^2}(\omega_0-\omega_p), \quad C_0=\sqrt{\frac{2\beta}{\gamma}}. \label{defa}
\end{equation}
In this case the sum in Eq.~(\ref{e8}) becomes exact if we assume an infinitely large ensemble of pulses giving \cite{r15}:
\begin{eqnarray}
a(\theta)&=& \theta_H C_0 dn\left[\frac{\theta}{\ell_H},\kappa \right],  \label{e9}\\
\theta_H&=&\frac{2K'}{\pi}C_0, \quad \ell_H= \frac{\pi}{2K'}\sqrt{\frac{D_2}{2(\omega_0-\omega_p)}}, \label{params}
\end{eqnarray}
where $K'=K(1-\kappa^2)$. It turns out that the spatial multiplex of sech pulses (\ref{e8}) is nothing else but an Elliptic soliton, and as such it is a soliton crystal equivalent to the elliptic-soliton solution (\ref{e11}) of the NLSE with a spatial repetition rate $\theta_0$ given by:
\begin{equation}
\theta_0=2K\ell_H, \label{dna}
\end{equation}
where $\ell_H$, the pulse width in the comb structure, has been defined in formula (\ref{params}). \\
The Elliptic solitons (\ref{e11}) or (\ref{e9}) form a multi-soliton complex that decay into an harmonic wavepacket when $\kappa \rightarrow 0$. When $\kappa \rightarrow 1$, both reduce to a single-pulse field. In the context of laser applications they exhibit interesting features among which the dependence of their amplitude on both the coupled resonance width and the pulse repetition rate. As it is apparent from formula (\ref{dna}), the pulse repetition rate is relevant in the sense that it determines the existence and stability of the soliton crystal in the ring resonator device: the smaller the pulse repetition rate the larger the pulse amplitude. Also, according to the above relations, the Elliptic soliton amplitude is inversely proportional to the cavity resonance width $\beta$ as defined in Eq. (\ref{defa}). Consequently the Elliptic soliton envelope should be widened as the amplitude grows, thereby reducing collisions between pulses in th soliton-comb structure. 

\section{\label{sec:3} Soliton crystal stability}
The stability of nonlinear signals in optical media is a key requirement for their processing, storage or transmission. While the issue has been investigated at length for single-pulse solutions (see e.g. ref.~\cite{r18}), recent interest in multiplexed solitons has shifted attention to the stability of these particular structures from both experimental~\cite{r16} and theoretical standpoints within the framework of the linear stability analysis~\cite{r1}. \\
In this section we wish to investigate the stability of soliton crystals, represented by Elliptic solitons as established in the previous section. In this purpose we consider small-amplitude noises in the ring microresonator propagating together with the Elliptic-soliton signal. For the sake of formal mathematical analysis, we shall deal with the Elliptic soliton (\ref{e11}) obtained as exact solution to the NLSE~(\ref{e2}). In the presence of small-amplitude noises this solution now becomes:
\begin{eqnarray}
\psi(\tau,\theta) =\{a(\theta) +[u(\theta) - v(\theta)]e^{i\omega \tau} + [u^*(\theta) + \nonumber \\ v^*(\theta)]e^{-i\omega \tau} \}e^{i \beta \tau},
\label{e12}
\end{eqnarray}
where $a(\theta)$ is the Elliptic-soliton envelope given by (\ref{e11}), while $u(\theta)$ and $v(\theta)$ are the spatial amplitudes of two complex noise fields having a common frequency $\omega$ \cite{r18}. Substituting Eq.~(\ref{e12}) in the NLSE~(\ref{e2}), and keeping only linear terms in $u(\theta)$ and $v(\theta)$, we find the following set of coupled Lam\'e \cite{sack} equations:
\begin{eqnarray}
\lbrack\frac{\partial^2}{\partial \varphi^2} + 2 \textbf{dn}^2(\varphi) + \varepsilon\rbrack v + \nu u &=& 0, \label{e13a} \\
\lbrack \frac{\partial^2}{\partial \varphi^2} + 6 \textbf{dn}^2(\varphi) + \varepsilon'\rbrack u + \nu v &=& 0, \label{e13b}
\end{eqnarray}
with $\varphi = \sqrt{-\frac{2 \beta}{\beta_2 (2-\kappa^2)}}\theta$, $\varepsilon = \varepsilon' = (2-\kappa^2)\beta$ and $\nu = (2-\kappa^2)\omega$. The set Eqs.~(\ref{e13a})-(\ref{e13b}) can be mapped onto a $2\times 2$-matrix linear eigenvalue problem with eigenfunctions forming a family of two-component complex vectors ($u$, $v$), whose eigenvalues are the couples of scalars ($\varepsilon$, $\varepsilon'$). In matrix form the eigenvalue problem reads:
\begin{eqnarray}
\textbf{H} \textbf{U} = \varepsilon(\kappa) \textbf{U},
\label{e14}
\end{eqnarray}
where
\[
\textbf{H}=
\begin{bmatrix}
    \frac{\partial^2}{\partial \varphi^2} + 2dn^2(\varphi)       & \nu \\
    \nu      &     \frac{\partial^2}{\partial \varphi^2} +6dn^2(\varphi) 
\end{bmatrix},
\quad \textbf{U}=
\begin{bmatrix}
    v   \\
    u      
\end{bmatrix}.
\]

We look for solutions to the above matrix eigenvalue problem, which are linear combinations of the exact eigenfunctions of equations (\ref{e13a}) and (\ref{e13b}) in the absence of coupling. To find these exact eigenfunctions we remark that for $\omega=0$, i.e. when the noise fields are in steady state, the set (\ref{e13a})-(\ref{e13b}) bursts into two independent Lam\'e equations: the first becomes a Lam\'e equation of first order and the second to a Lam\'e equation of second order. Instructively, an $n^{th}$ order Lam\'e eigenvalue problem has the general form \cite{r15,kabaz,dikpsc}:
\begin{equation}
-\lbrack \frac{\partial^2}{\partial z^2} +n(n+1)\kappa^2sn^2(z)\rbrack \psi(z)= \lambda(\kappa)\psi(z), 
\end{equation}
where $sn$ is a Jacobi elliptic function and $n$ a positive integer. Considering the discrete spectrum of this eigenvalue equation, it is known that for a given value of $n$ the Lam\'e equation possesses exactly $2n+1$ boundstate solutions. For Eqs. (\ref{e13a})-(\ref{e13b}) this leads to the eigenfunctions listed in tables \ref{tab:table1} and \ref{tab:table2}, together with their corresponding eigenvalues.  

\begin{table}
\caption{\label{tab:table1} Eigenvalues $\varepsilon(\kappa)$ and eigenfunctions $v(\varphi)$ of the first-order Lam\'e equation~(\ref{e13a}), when $\omega=0$. $v_0^{(i)}$ are normalization constants.}
\begin{ruledtabular}
\begin{tabular}{lcr}
Eigenvalues & Eigenfunctions \\
\hline
 $\varepsilon(\kappa)=(1+\kappa^2)$ & $v^{(1)}(\kappa)=v_{0}^{(1)}sn(\varphi)$\\
 $\varepsilon(\kappa)=1$ & $v^{(2)}(\kappa)=v_{0}^{(2)}cn(\varphi)$  \\
 $\varepsilon(\kappa)=\kappa^2$ &  $v^{(3)}(\kappa)=v_{0}^{(3)}dn(\varphi)$\\
\end{tabular}
\end{ruledtabular}
\end{table}

\begin{table}
\caption{\label{tab:table2} Eigenvalues $\varepsilon'(\kappa)$ and eigenfunctions $u(\varphi)$ of the second-order Lam\'e equation~(\ref{e13b}), when $\omega=0$. $\kappa_1^2=1-\kappa^2$, $u_0^{(i)}$ are normalization constants.}
\begin{ruledtabular}
\begin{tabular}{lcr}
Eigenvalues & Eigenfunctions\\
\hline
 $\varepsilon'(\kappa)=(4+\kappa^2)$ & $u^{(1)}(\kappa)=u_{0}^{(1)}sn(\varphi)cn(\varphi)$\\
 $\varepsilon'(\kappa)=(1+4\kappa^2)$ & $u^{(2)}(\kappa)=u_{0}^{(2)}sn(\varphi)dn(\varphi)$  \\
 $\varepsilon'(\kappa)=(1+\kappa^2)$ & $u^{(3)}(\kappa)=u_{0}^{(3)}cn(\varphi)dn(\varphi)$ \\
 $\varepsilon'(\kappa)=2[(1+\kappa^2)\mp \sqrt{1-\kappa^2\kappa_1^2}]$ & $u^{(4,5)}(\kappa)=u_{0}^{(4,5)}sn^2(\varphi)-u_{0}^{4,5}$\\
    & $\times\frac{(1+\kappa^2)\pm \sqrt{1-\kappa^2\kappa_1^2}}{3k^2}$ \\
\end{tabular}
\end{ruledtabular}
\end{table}

\par
\begin{figure}[h]
\includegraphics[width=1.6in,height=1.3in]{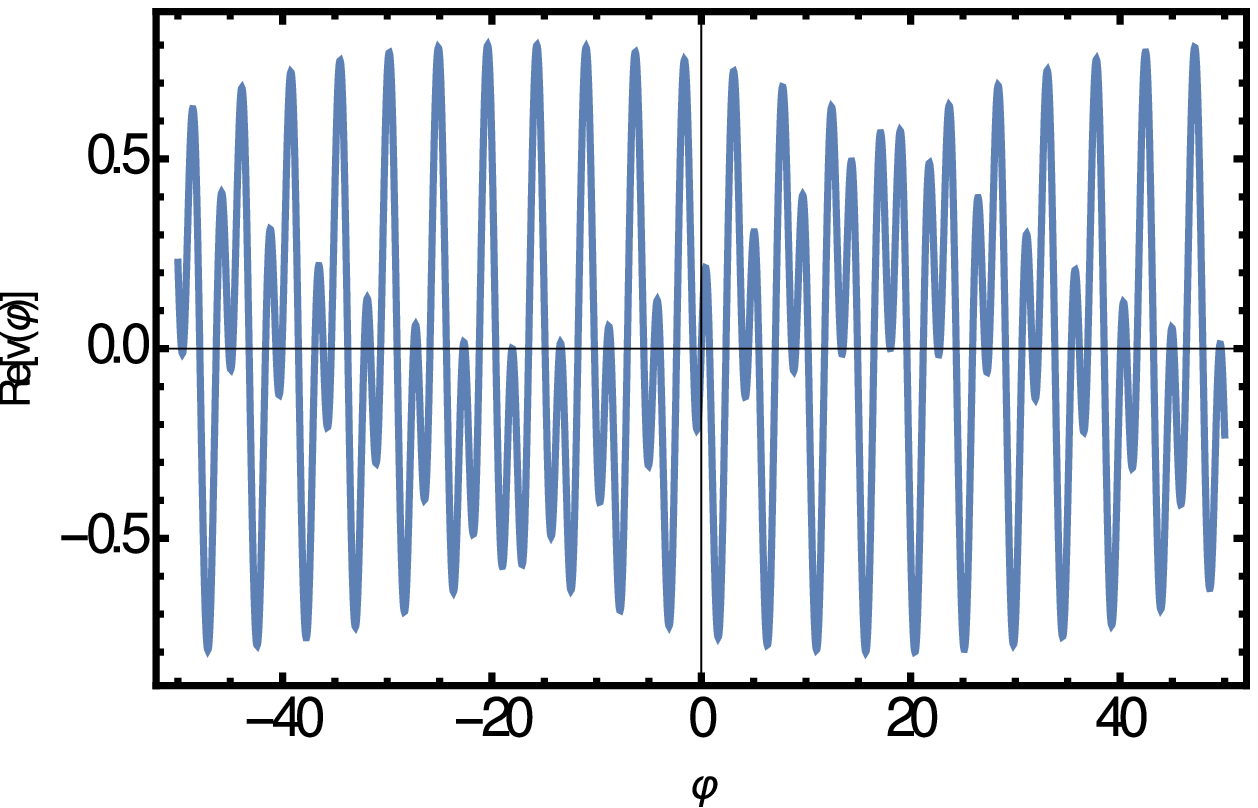}
\includegraphics[width=1.6in,height=1.3in]{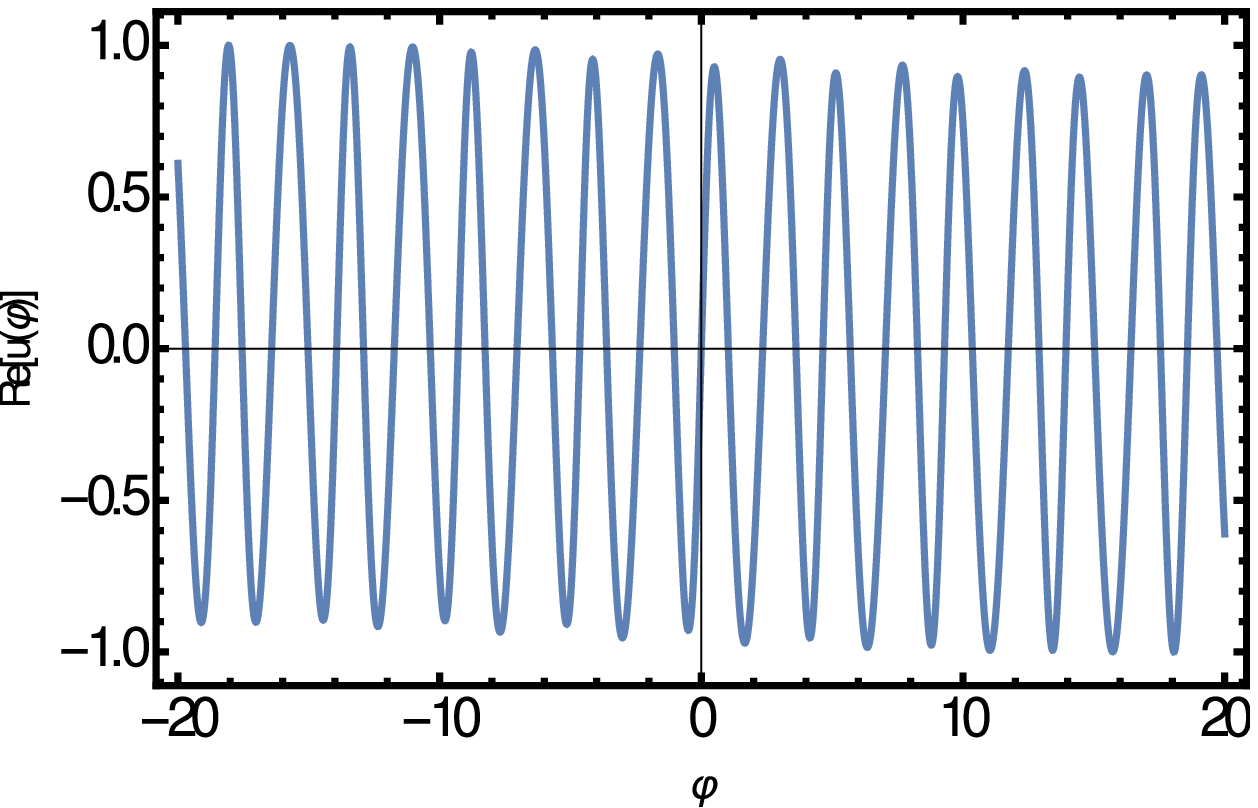}
\includegraphics[width=1.6in,height=1.3in]{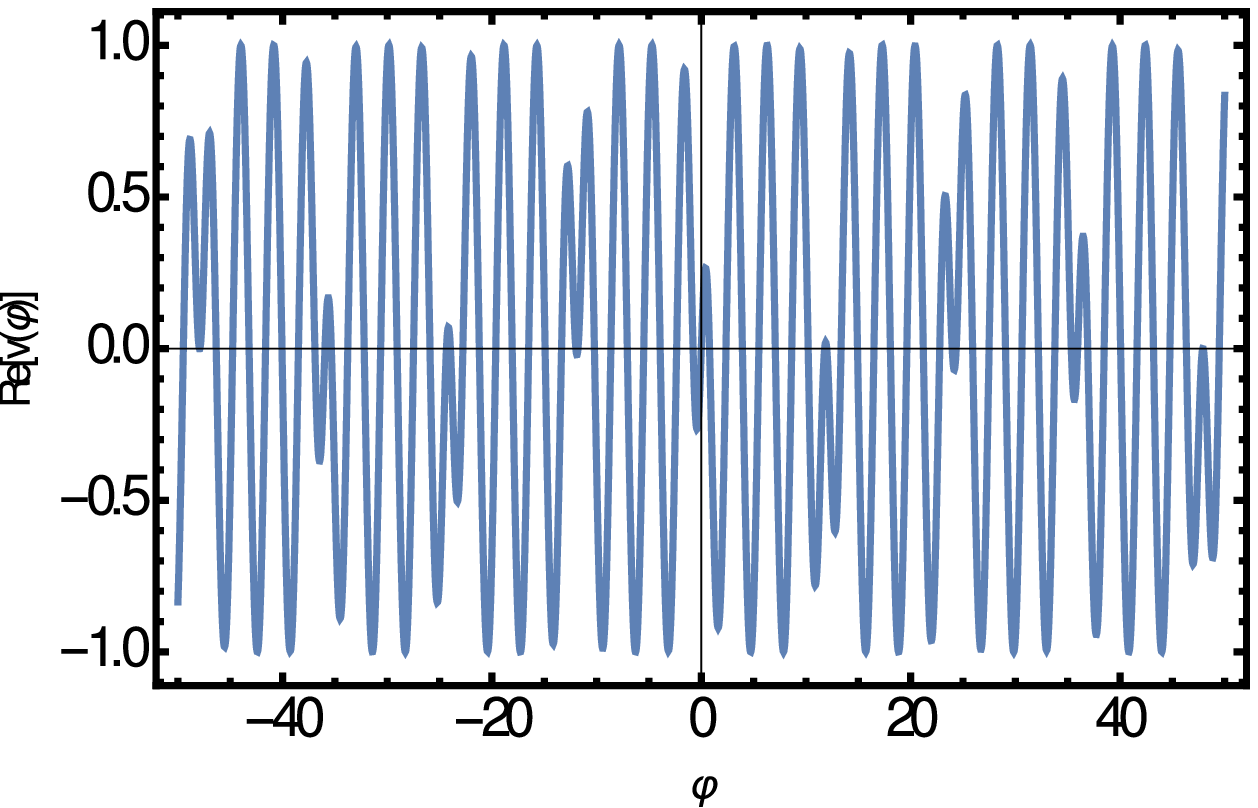}
\includegraphics[width=1.6in,height=1.3in]{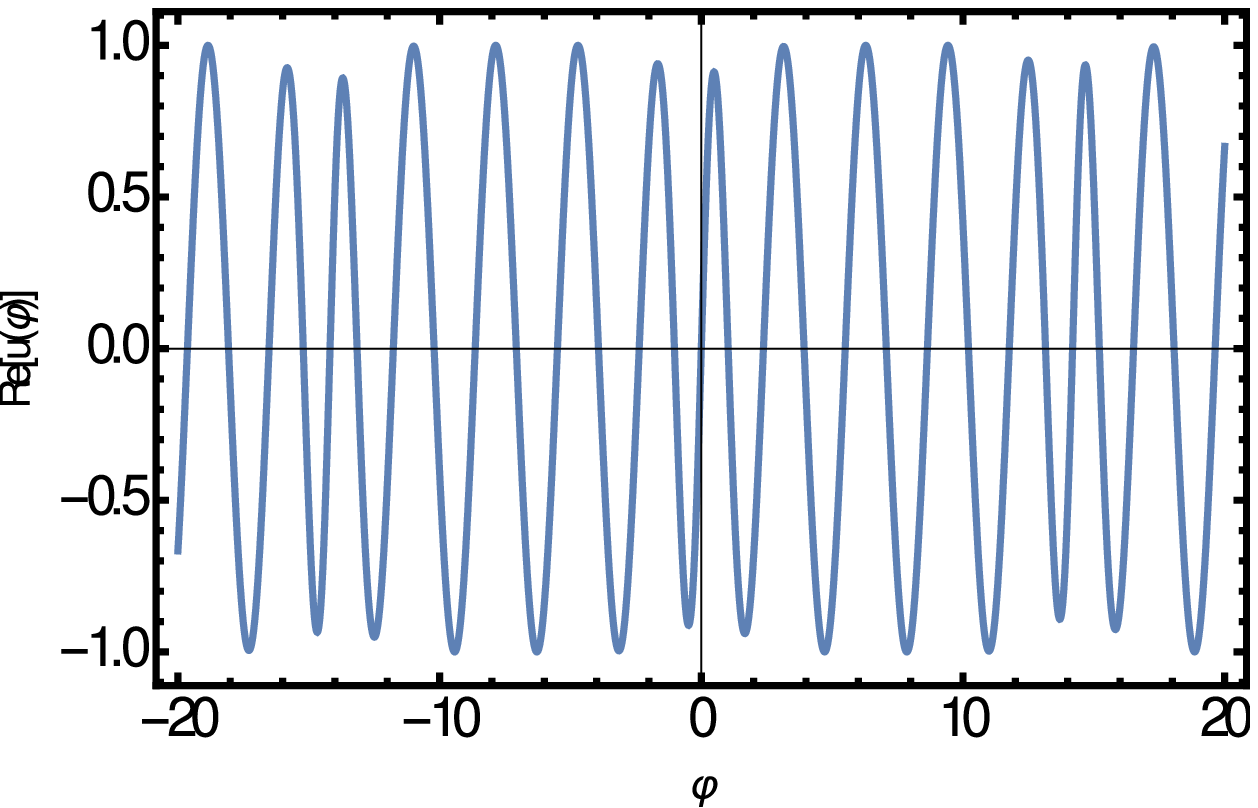}
\caption{\label{f2}
(Color online) Real parts of the internal modes v($\varphi$) (left) and u($\varphi$) (right) given in (\ref{e17a}) and (\ref{e17b}). Top: $\kappa=0.8$, bottom: $\kappa=97$.} 
\end{figure}

\begin{figure}[h]
\includegraphics[width=1.6in,height=1.3in]{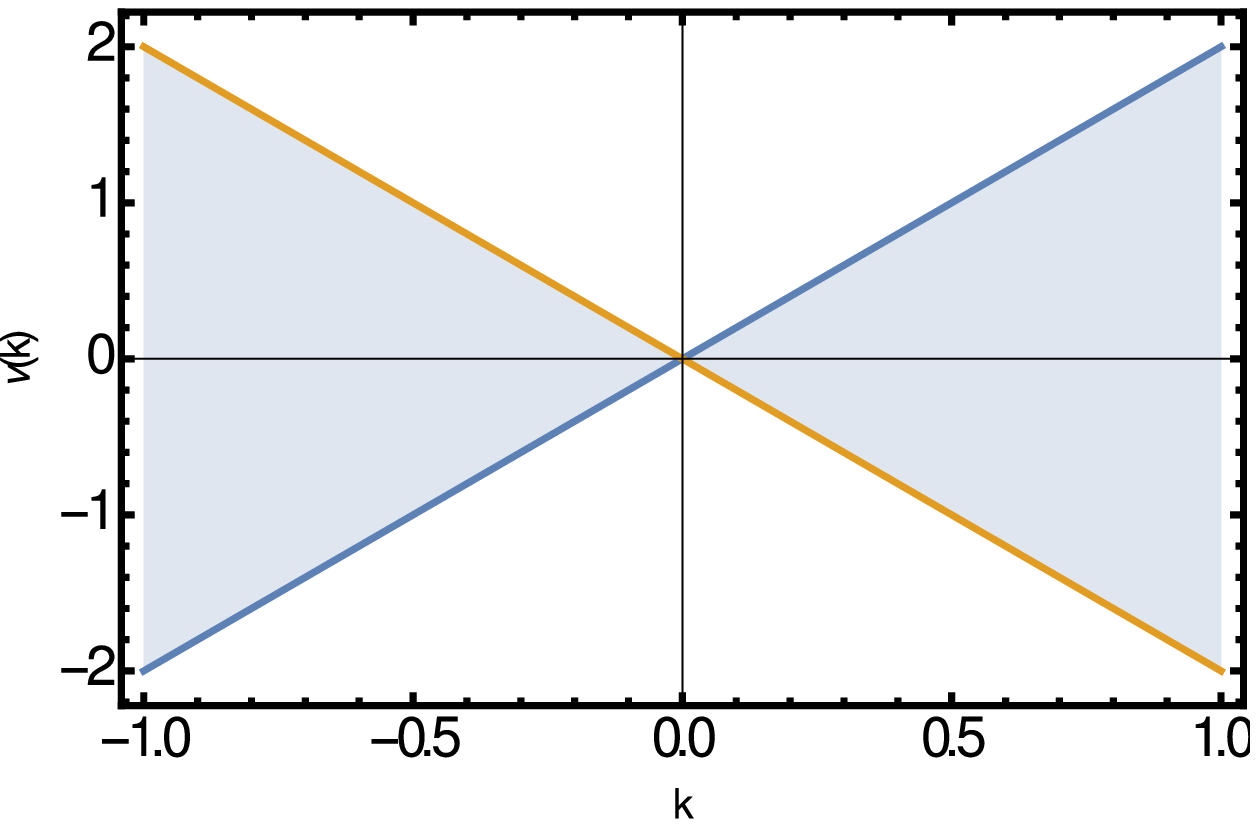}
\includegraphics[width=1.6in,height=1.3in]{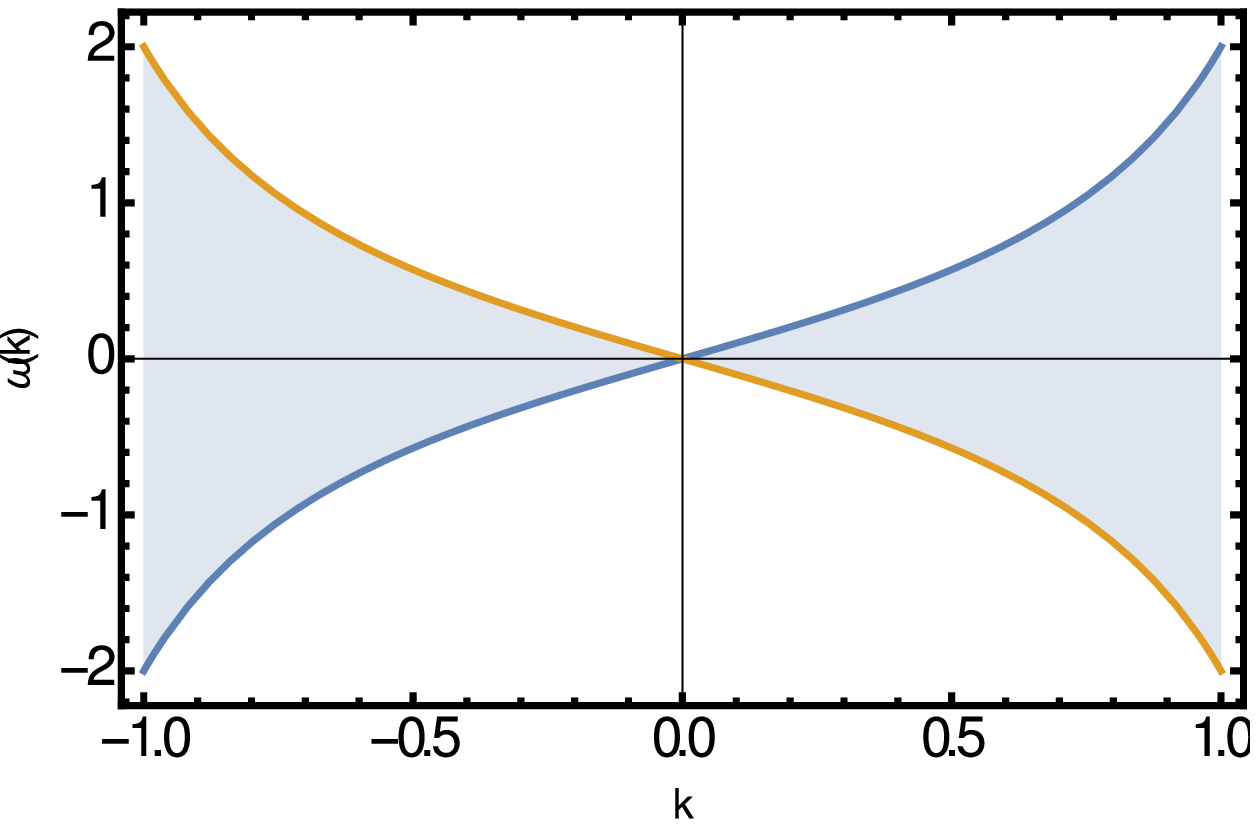}
\caption{(Color online) Plots of the internal-mode eigenvalues $\nu$ (right) and $\omega$ (left) given by (\ref{val2}),
as a function of the Jacobi elliptic modulus $\kappa$. In the two plots the shaded region (blue region in color) denotes the stability domain.} 
\label{f3}
\end{figure}

 When $\omega\neq 0$, solutions to the coupled eigenvalue equations (\ref{e13a}) and (\ref{e13b}) are vectors $\lbrack u(\varphi)$, $v(\varphi)\rbrack $ whose components are orthogonal linear combinations of the two-component basis vector $u_0(\varphi)$, $v_0(\varphi)$, which are exact eigenmodes given in tables \ref{tab:table1} and \ref{tab:table2}. Two new basis vectors ensuing from the requirement of an orthonormalized basis for the two coupled Lam\'e equations (\ref{e13a})-(\ref{e13b}) are:  
\[
\begin{bmatrix}
    a_1 \\
    1
\end{bmatrix}cn(\varphi)e^{i \eta} +
\begin{bmatrix}
    0 \\
    b_1
\end{bmatrix}sn(\varphi)dn(\varphi)e^{i \eta}
,
\]

\[
\begin{bmatrix}
    a_2 \\
    1
\end{bmatrix}sn(\varphi)e^{i \eta} +
\begin{bmatrix}
    0 \\
    b_2
\end{bmatrix}cn(\varphi)dn(\varphi)e^{i \eta},
\]
where the functions $f(x,\kappa)=\{cn(x), sn(x)\}$ and $g_i(x,\kappa)=\{sn(x)dn(x), cn(x)dn(x)\}$ should be mutually orthogonal i.e.
\begin{eqnarray}
\int_{-K}^{K} f_i(x,\kappa) g_i(x,\kappa) dx = 0.
\label{e15}
\end{eqnarray}
In this new two-component basis we obtain the following solutions:
\begin{itemize}
 \item On the basis $\lbrack cn(x), sn(x)dn(x)\rbrack$:
 \begin{eqnarray}
v(\varphi)&=& \mp \frac{i\kappa}{\sqrt{1-\kappa^2}}cn(\varphi,\kappa)e^{2i \sqrt{1-\kappa^2} \varphi}, \label{e16a}
\\
u(\varphi)&=& \left[cn(\varphi,\kappa) + \frac{i}{\sqrt{1-\kappa^2}} sn(\varphi,\kappa)dn(\varphi,\kappa)\right]
e^{2i \sqrt{1-\kappa^2} \varphi}, \nonumber  \\
\label{e16b}
\end{eqnarray}
with spectral parameters:
\begin{equation}
\nu(\kappa)= \mp 2i\kappa \sqrt{1-\kappa^2}, \quad \varepsilon(\kappa)=1 - 2\kappa^2. \label{val1}
\end{equation}
\item On the basis $\lbrack sn(x), cn(x)dn(x)\rbrack$:
 \begin{eqnarray}
v(\varphi)&=& \pm sn(\varphi,\kappa)e^{2i \varphi}, \label{e17a}
\\
u(\varphi)&=& \left[sn(\varphi,\kappa) \mp i cn(\varphi,\kappa)dn(\varphi,\kappa)\right]
e^{2i \varphi},
\label{e17b}
\end{eqnarray}
for which the spectral parameters are:
\begin{equation}
\nu(\kappa)= \pm 2\kappa, \quad \varepsilon(\kappa)=1 + \kappa^2. \label{val2}
\end{equation}
According to Eq. (\ref{val1}) eigenvalues of the internal modes (\ref{e16a})-(\ref{e16b}) are purely imaginary, therefore these modes will either decay or grow and hence are unstable. We refer to the internal mode with a growing amplitude as "growing modes", whereas the internal mode with decreasing amplitude is referred to as "decaying modes". In fact their unstability affects the Elliptic-soliton stability, indeed the ansatz formula (\ref{e12}) clearly shows that the exponential growth or decay of internal modes are the result of a combination with the exponential modulation of the Elliptic soliton envelope $a(\varphi)$. \\
On the contrary eigenvalues of the pair of internal modes (\ref{e17a})-(\ref{e17b}) are real, as evidenced by formula (\ref{val2}). Hence they are stable boundstates, co-existing with the Elliptic soliton but distinct from the Elliptic soliton by their amplitudes and frequencies. They can be regarded as long-term (i.e. nonlinear) periodic oscillations in the Elliptic-soliton background. Their amplitudes are represented in fig.~\ref{f2}, while their eigenvalues are sketched in fig.~\ref{f3}.
\end{itemize}

\section{\label{sec:4} Dynamics of Elliptic solitons in ring microresonators: Collective coordinate approach}

\subsection{\label{subsec:41}Collective-Coordinate equations for the general perturbed NLSE}

Unlike the NLSE for which the Elliptic-soliton solution has been obtained in sec. \ref{sec:2} and shown to be equivalent to a pulse-shaped soliton comb, the LLE Eq.~(\ref{e1}) is not exactly integrable. Nevertheless an approximate solution for this last equation can be obtained by means of a perturbation theory, such as the collective-coordinate method which is more explicitely an adiabatic perturbation theory. While this method has been widely utilized for perturbed NLSEs in the context of single-pulse soliton \cite{r20}, the possible existence of multi-pulse solutions and particularly soliton-crystal solutions to the same equations recently motivated its extension \cite{r19} to these particular physical contexts. To start we shall construct the collective-coordinate theory for the Elliptic-soliton solution to the NLSE with an arbitrary perturbation, and next apply the theory to the LLE. \\
Consider the perturbed cubic NLSE:
\begin{eqnarray}
i\frac{\partial \psi}{\partial t}+\frac{1}{2}\frac{\partial^2 \psi}{\partial \vartheta^2}+\gamma\rvert \psi
\rvert^2\psi(t,\vartheta)= i\epsilon Q(\psi,\psi^{(n,m)}),
\label{e18}
\end{eqnarray}
where $\epsilon Q(\psi,\psi^{(n,m)})$ represents the perturbation and $\psi^{(n,m)}$ refer to the $n^{th}$ and $m^{th}$ derivatives of
$\psi(t,\vartheta)$ with respect to $t$ and $\vartheta$ respectively. Let the Elliptic-soliton solution to Eq. (\ref{e18}) be of a general form:
\begin{eqnarray}
\psi(t,\vartheta)=\eta(t)dn\{\eta(t)[\theta - q(t)]\}exp[i\phi(t)-\delta(t)\vartheta], \nonumber \\
\label{e19}
\end{eqnarray}
where $\eta(t)$ is the amplitude, $q(t)$ is the position, $\phi(t)$ is the phase of internal oscillations and $\delta(t)$ is the phase velocity of the Elliptic soliton. Our aim is to determine the set of differential equations governing time evolutions of these four collective-coordinate variables. Within the framework of the Lagrangian formalism \cite{r19}, we define the Lagrangian density associated to the perturbed Eq.~(\ref{e18}) as:
\begin{eqnarray}
l_d=\frac{i}{2}(\psi\psi^*_t-\psi_t\psi^*)-\frac{1}{2}(\vert\psi\vert^4-\vert\psi_t\vert^2) \nonumber \\ + i(\epsilon Q\psi^* - \epsilon Q^*\psi),
\label{e20}
\end{eqnarray}
where subscripts refer to derivatives with respect to the variable. The Lagrangian $L$ can be computed via the formula:
\begin{eqnarray}
L=\int^{K}_{- K}l_d\quad d\vartheta,
\label{e21}
\end{eqnarray}
which after substitution of Eq. (\ref{e19}) and integration yields:
\begin{eqnarray}
L=(\phi_t - \delta_t q)\eta E -\frac{4}{3}(2-\kappa^2) \eta^3 E
 +\eta [(2-\kappa^2)\eta^2 \nonumber \\ + \delta^2]E +i\int^{K}_{- K} (\epsilon Q\psi^*-
\epsilon Q^*\psi) d\vartheta,  \hspace{.3in}
\label{e22}
\end{eqnarray}
where $E$ is the elliptic integral of second kind. Applying the Lagrangian formalism with respect to the four collective-coordinate variables on $L$ \cite{r19}, we obtain the following set of coupled first-order time ordinary differential Equations:
\begin{eqnarray}
\eta_t&=& \frac{1}{E}Re \int^{K}_{-K} \epsilon Q(\psi)\psi\, d\vartheta, \label{e23a} \\
\delta_t &=& -\frac{\kappa^2}{E}Im \int^{K}_{-K}sn[\eta(\vartheta-q)] \epsilon Q(\psi)\psi\,d\vartheta, \label{e23b} \\
q_t&=&-\delta + \frac{1}{\eta^2 E}Re \int^{K}_{-K} (\vartheta-q)\epsilon Q(\psi)\psi\,d\vartheta, \label{e23c} \\
\phi_t &=& \frac{1}{E}Im \int^{K}_{-K} \epsilon Q(\psi)\{\frac{1}{\eta}\psi^* -\kappa^2 
(\vartheta -q) sn[\eta(\vartheta -q)]\psi^*_{cn}\}d\vartheta 
\nonumber \\ &+& \frac{1}{2}[(2-\kappa^2)\eta^2 -\delta^2] + q\delta_t. \label{e23d}
\label{e23}
\end{eqnarray}
In the last set we defined:
\begin{equation}
\psi_{cn}(t,\vartheta)=\eta(t)cn\{\eta(t)[\theta - q(t)]\}exp[i\phi(t)-\delta(t)\vartheta], \label{defb}
\end{equation}
while "Re" and "Im" stand for real and imaginary parts respectively. 

\subsection{\label{subsec:42}Collective-Coordinate equations for the LLE}
By setting $t=\frac{\gamma}{2}\tau$ and $\vartheta=\sqrt{\frac{\gamma}{2 \vert\beta_2\vert}}\theta$, the LLE Eq.~(\ref{e1}) can be expressed as the perturbed NLSE Eq.~(\ref{e18}) with:
\begin{equation}
\epsilon Q(\psi)=f -\frac{2}{\gamma}(\alpha_1+i\alpha_2)\psi(t,\vartheta),
 \label{e24}
\end{equation}
where $f=\frac{2}{\gamma}F$. Replacing Eq.~(\ref{e24}) in the set Eqs.~(\ref{e23a})-(\ref{e23d}) we obtain:
\begin{eqnarray}
\eta_t&=& -2\alpha_r\eta +\frac{2f}{E}cos(\delta q - \phi) \int^{K}_{0}dn(\vartheta )cos\left(\frac{\delta}{\eta}\vartheta\right) d\vartheta, \nonumber \\
\label{e25a} \\
\delta_t&=& -\frac{2\kappa^2}{E}fcos(\delta q - \phi) \int^{K}_{0}dn(\vartheta )
sn(\vartheta ) sin\left(\frac{\delta}{\eta}\vartheta'\right) d\vartheta, \nonumber \\
\label{e25b} \\
q_t&=& - \delta -\frac{2f}{\eta^3 E}sin(\delta q - \phi) \int^{K}_{0}\vartheta dn(\vartheta )sin\left(\frac{\delta}{\eta}\vartheta\right) d\vartheta, \nonumber \\ 
\label{e25c} \\
\phi_t &=& -2\alpha_i - \frac{\alpha_i}{E}[E - (1-\kappa^2)K] + \frac{2f}{\eta E}
sin(\delta q - \phi) \nonumber \\
&\times& \int^{K}_{0} dn(\vartheta )cos\left(\frac{\delta}{\eta}\vartheta'\right) d\vartheta- \frac{2\kappa^2}{\eta E}fcos
(\delta q - \phi) \nonumber \\
&\times& \int^{K}_{0}cn(\vartheta )sn(\vartheta ) sin\left(\frac{\delta}{\eta}\vartheta\right) d\vartheta \nonumber \\
&+&  \frac{1}{2}[(2-\kappa^2)\eta^2 -\delta^2] +  q\delta_t,
\label{e25d}
\end{eqnarray}
where we defined $\alpha_r= \frac{2 \alpha_1}{\gamma}$, $\alpha_i= \frac{2 \alpha_2}{\gamma}$. In the next section we present numerical simulations of the collective-coordinate equations (\ref{e25a})-(\ref{e25d}), and explore some important aspects of the system dynamics including the phase portraits and the energy of the Elliptic soliton for some values of the perturbation parameters.

\section{\label{sec:5}Numerical results}
\begin{figure}[h]
\includegraphics[width=1.6in,height=1.3in]{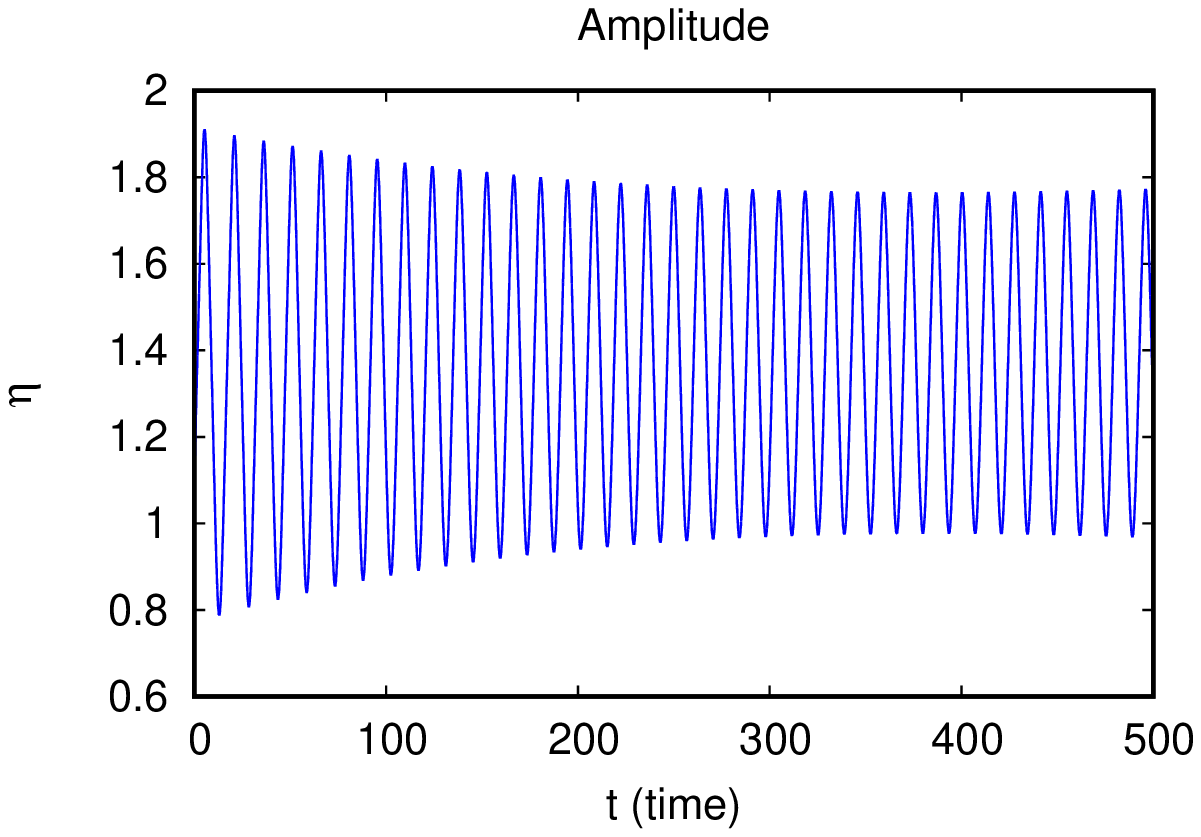} 
\includegraphics[width=1.6in,height=1.3in]{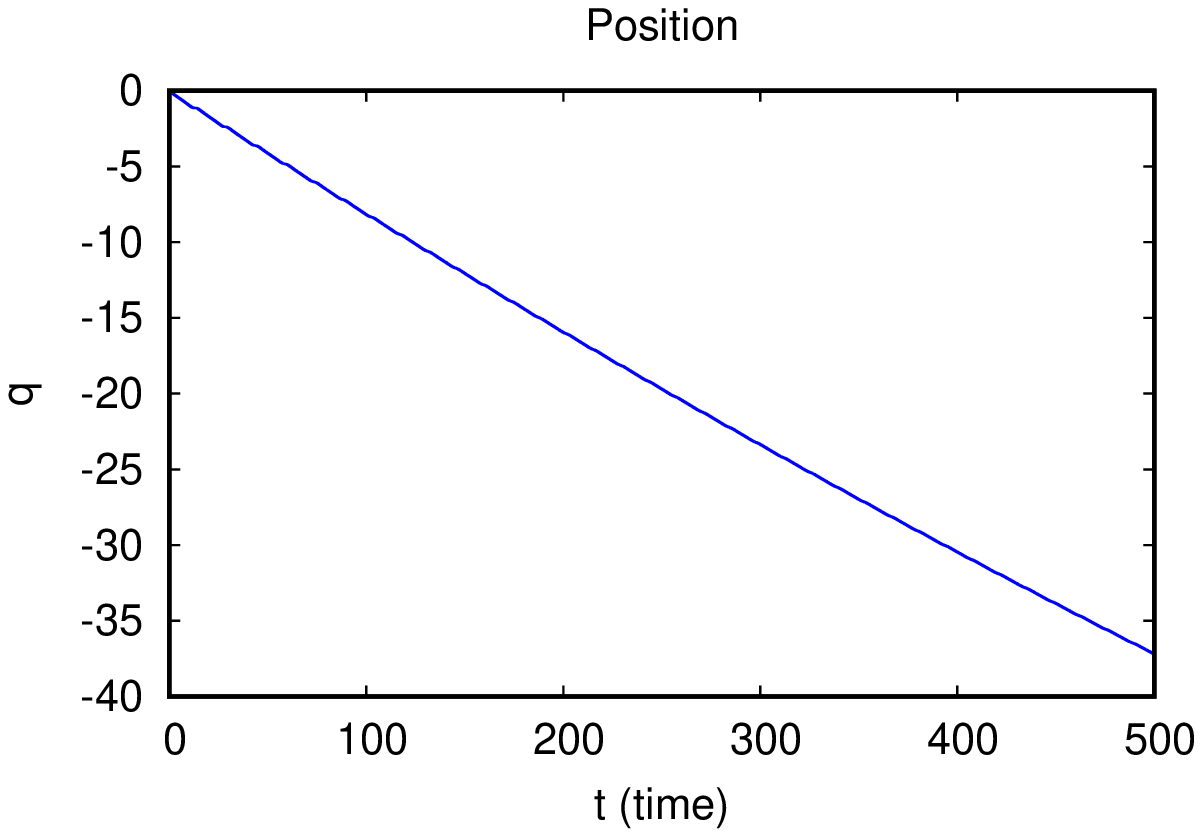}
\includegraphics[width=1.6in,height=1.3in]{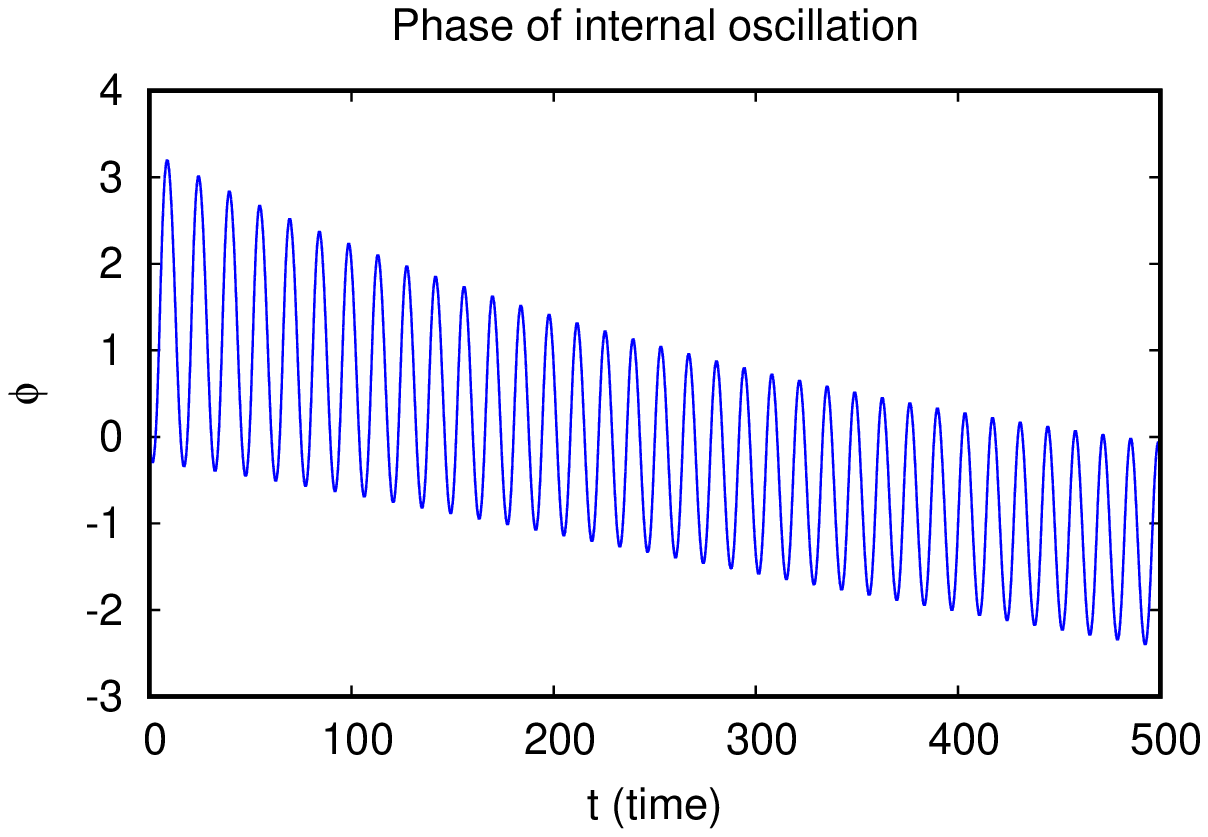}
\includegraphics[width=1.6in,height=1.3in]{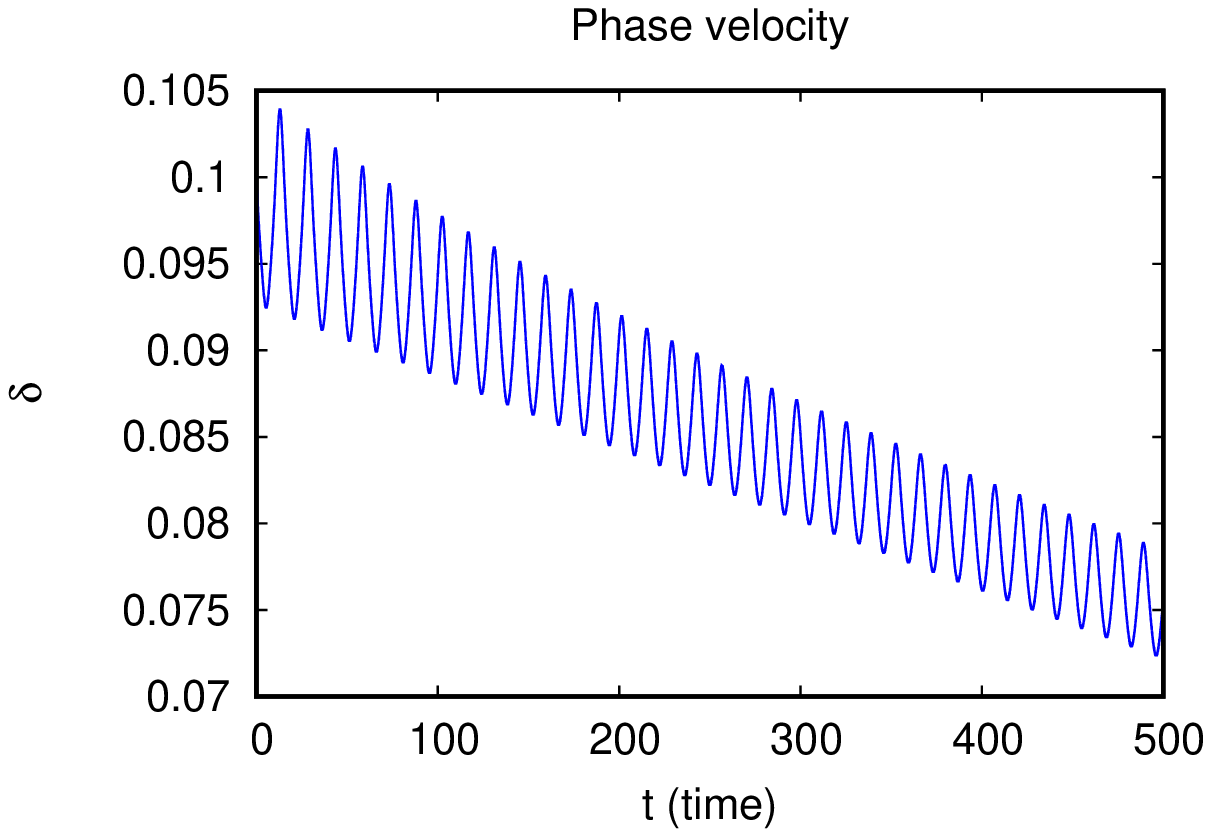}
\caption{\label{f4}
(Color online) Time evolutions of the Elliptic-soliton parameters $\eta(t)$, $q(t)$, $\phi(t)$ and $\delta(t)$ for $\alpha_1=0.001$ and $\alpha_2=0.5$.} 
\end{figure}

The collective-coordinate equations (\ref{e25a})-(\ref{e25d}) were solved numerically using a sixth-order Runge-Kutta scheme~\cite{luth,r33}, in conjunction with a $3/8$ Simpson rule for the integrals. The Jacobi elliptic functions $dn$, $sn$ and $cn$ were generated numerically by employing an algorithm proposed in~\cite{r23}. In all our simulations we considered $\kappa=0.97$, $\gamma=1$ and $F=0.01$. \\
Fig.~\ref{f4} shows time series of the amplitude $\eta(t)$, the position $q(t)$, the phase of internal oscillations $\phi(t)$ and the phase velocity $\delta(t)$ of the Elliptic soliton, obtained numerically for $\alpha_1=0.001$ and $\alpha_2=0.5$.  

\begin{figure}[h]
\includegraphics[width=1.51in,height=1.3in]{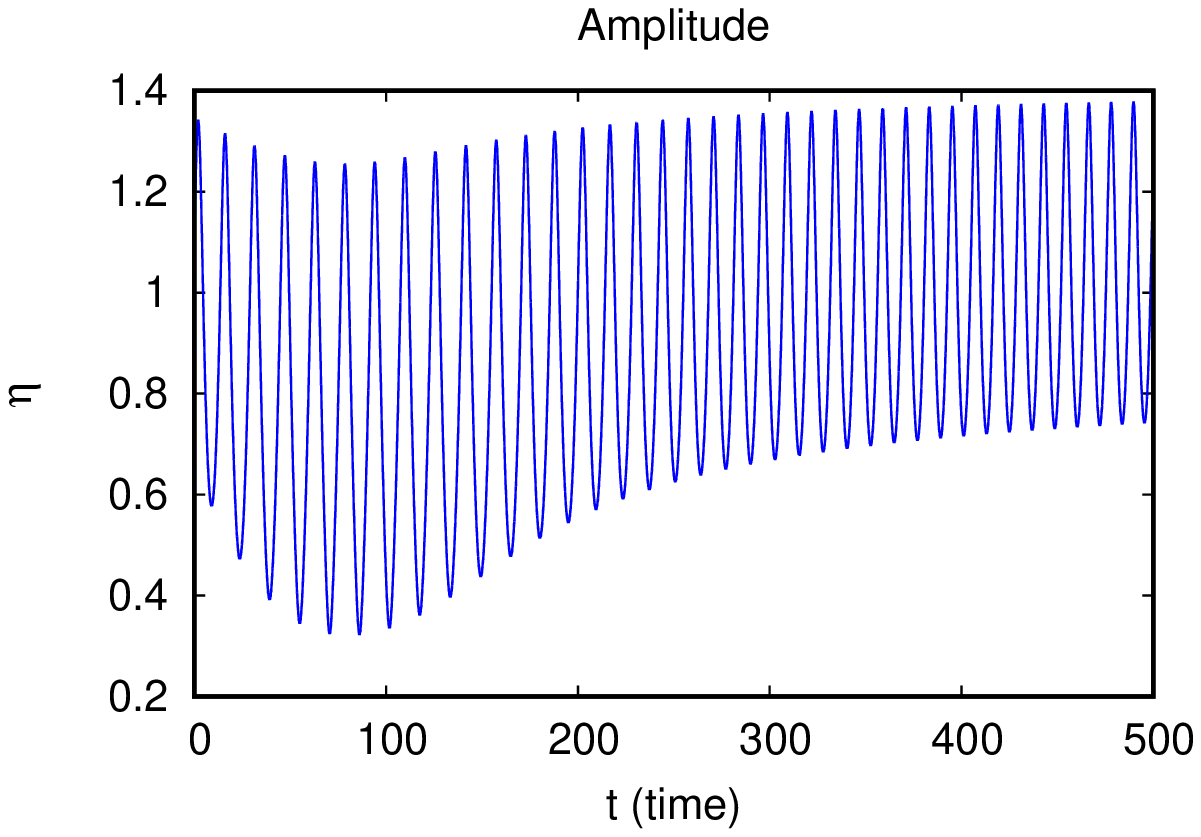}(a)
\includegraphics[width=1.51in,height=1.3in]{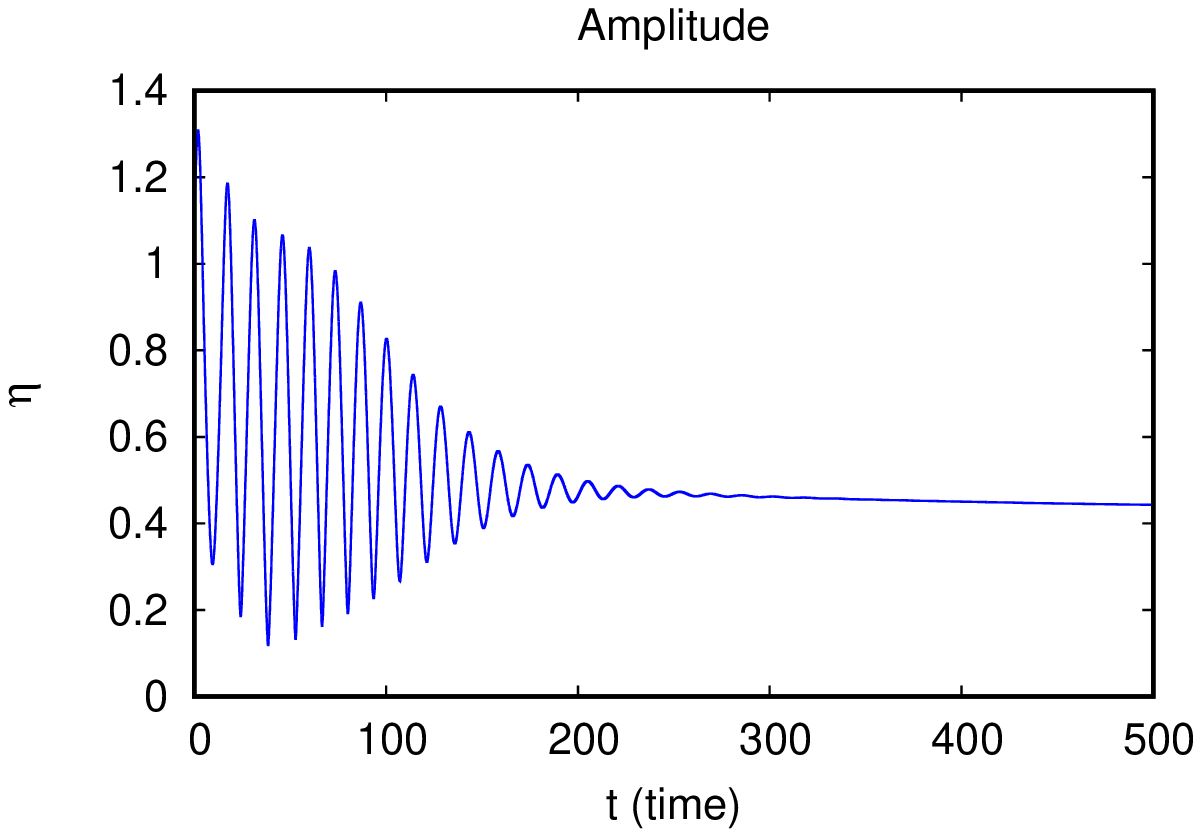}(b)
\includegraphics[width=1.51in,height=1.3in]{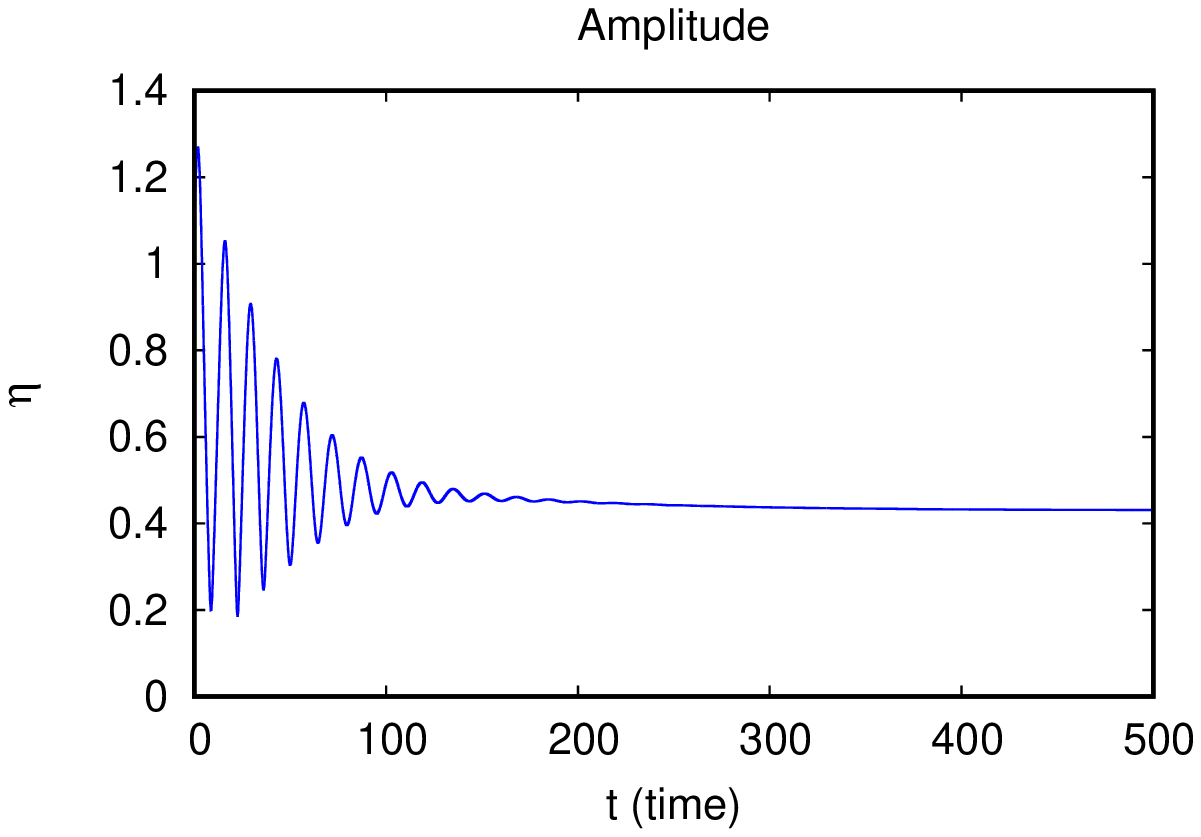}(c)
\includegraphics[width=1.51in,height=1.3in]{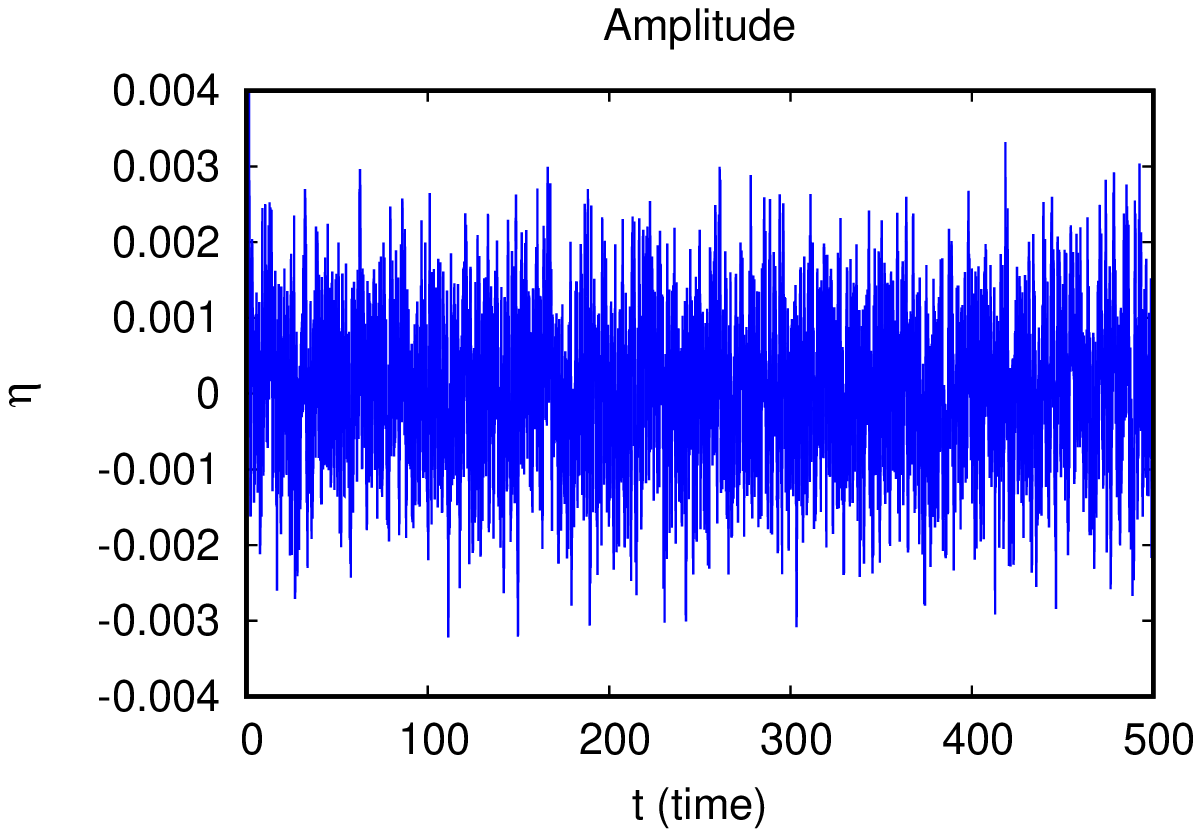}(d)
\caption{\label{f5}
(Color online) Time evolutions of the Elliptic-soliton amplitude $\eta(t)$ for $\alpha_2=0.01$. From (a) to (d): $\alpha_1=0.001$, $\alpha_1=0.005$ $\alpha_1=0.01$, $\alpha_1=1$.} 
\end{figure}

\begin{figure}[h]
\includegraphics[width=1.51in,height=1.3in]{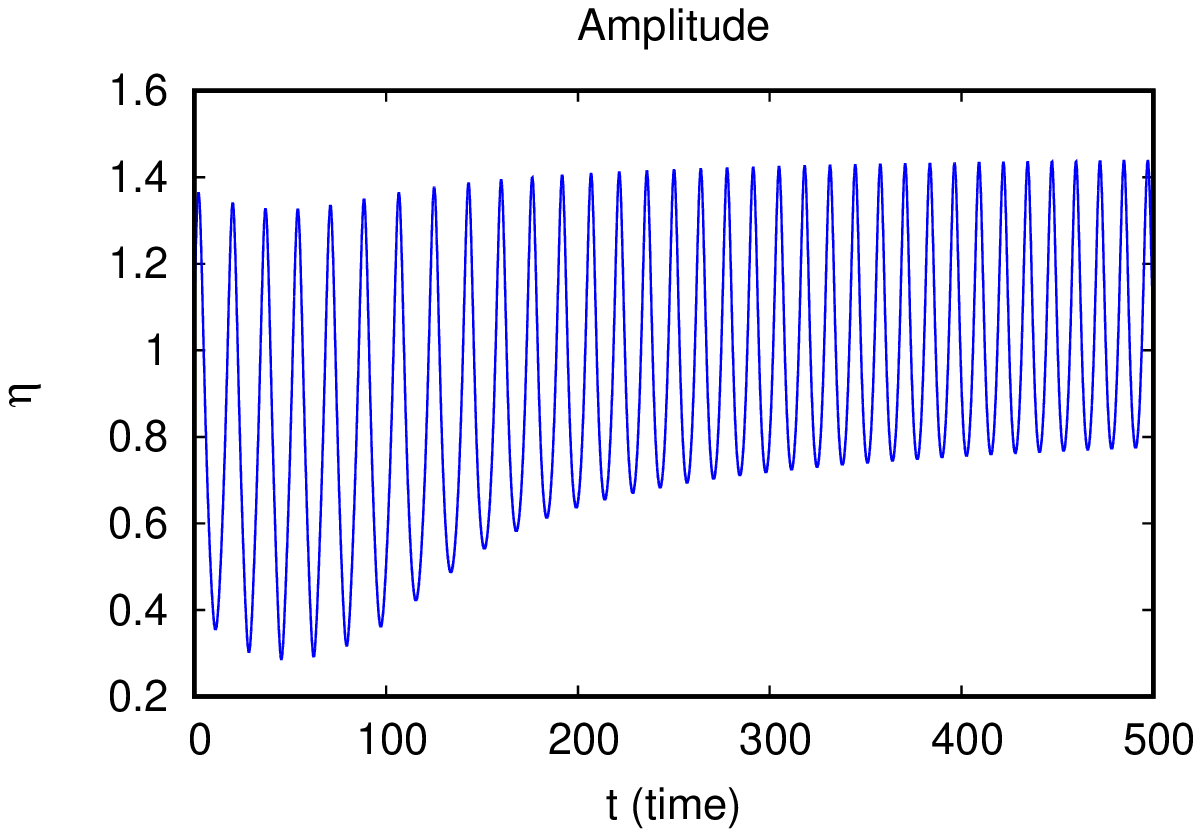}(a)
\includegraphics[width=1.51in,height=1.3in]{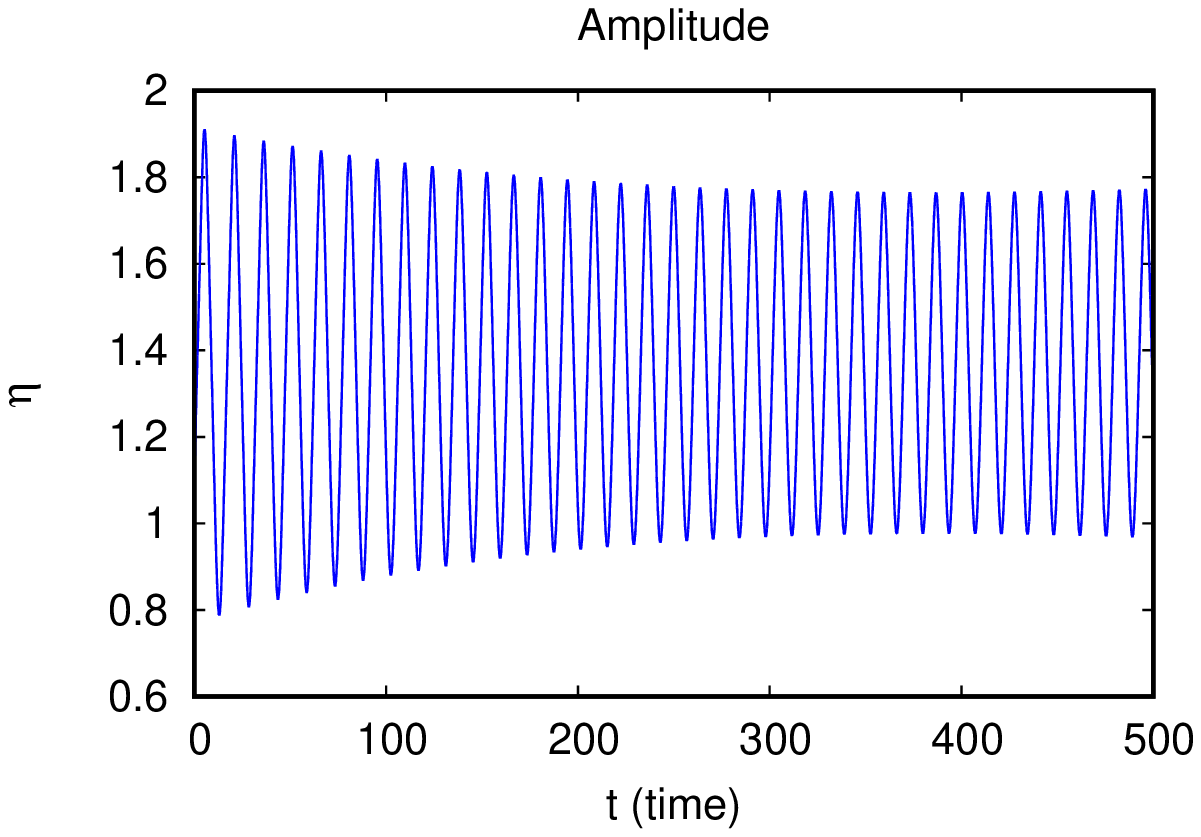}(b)
\includegraphics[width=1.51in,height=1.3in]{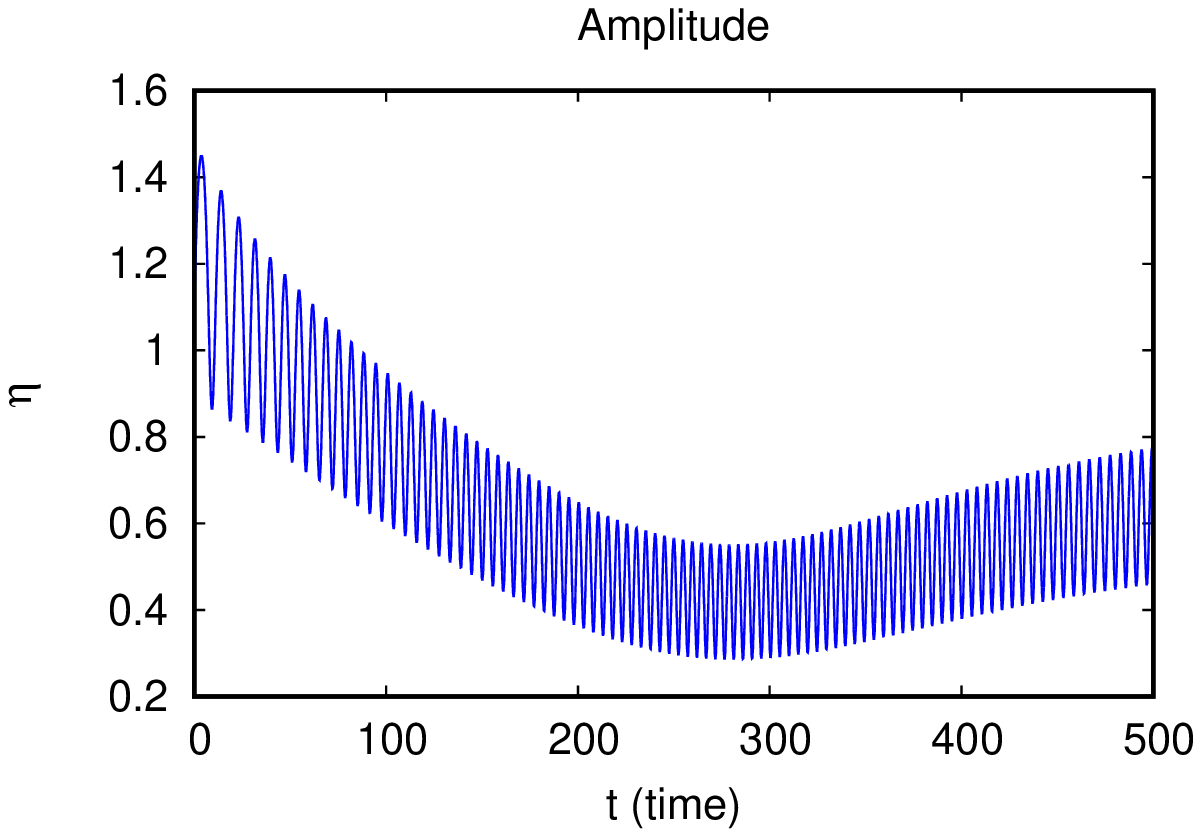}(c)
\includegraphics[width=1.51in,height=1.3in]{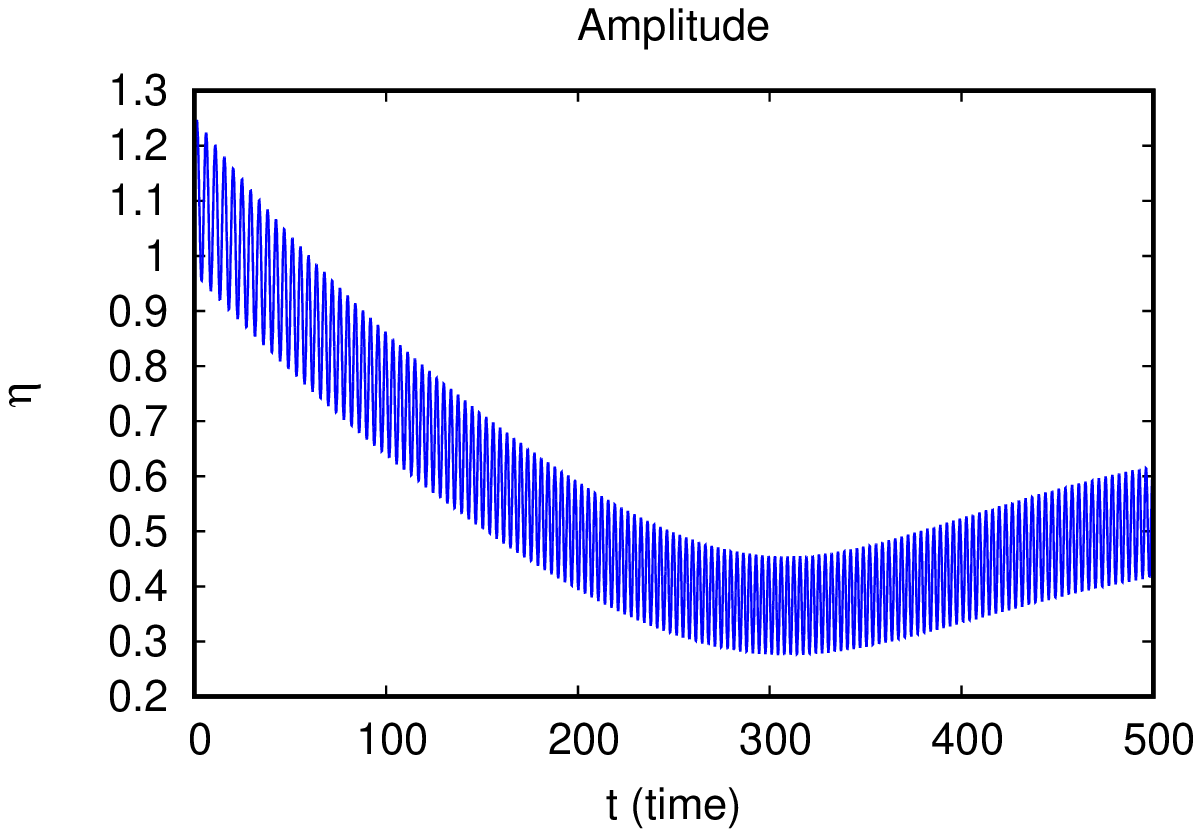}(d)
\caption{\label{f6}
(Color online) Time evolutions of the Elliptic-soliton amplitude $\eta(t)$, for $\alpha_1=0.001$. From (a) to (d): $\alpha_2=0.005$, $\alpha_2=0.5$, $\alpha_2=1$, $\alpha_2=1$.} 
\end{figure}

A main dominant behavior of characteristic parameters of the Elliptic soliton emerging from the figures is a damped oscillation, except the position which is only very slightly oscillating and suggests more a linear acceleration. The time series of the amplitude seems more appealing and informative about the Elliptic soliton stability, so we followed the evolution of this specific parameter for different combinations of the two perturbation coefficients $\alpha_1$ and $\alpha_2$. In particular fig.~\ref{f5} represents the time evolution of $\eta(t)$ for a fixed value of $\alpha_2$ but varying $\alpha_1$, whereas in fig. \ref{f6} $\alpha_1$ is kept fix but $\alpha_2$ is varied. If the exponentially damped oscillating behavior remains most dominant, the two figures clearly suggest that the linear loss coefficient $\alpha_1$ effectively controls the Elliptic soliton stability while an increase in $\alpha_2$ enhances the oscillatory feature of its amplitude. 

\begin{figure}[h]
\includegraphics[width=1.51in,height=1.3in]{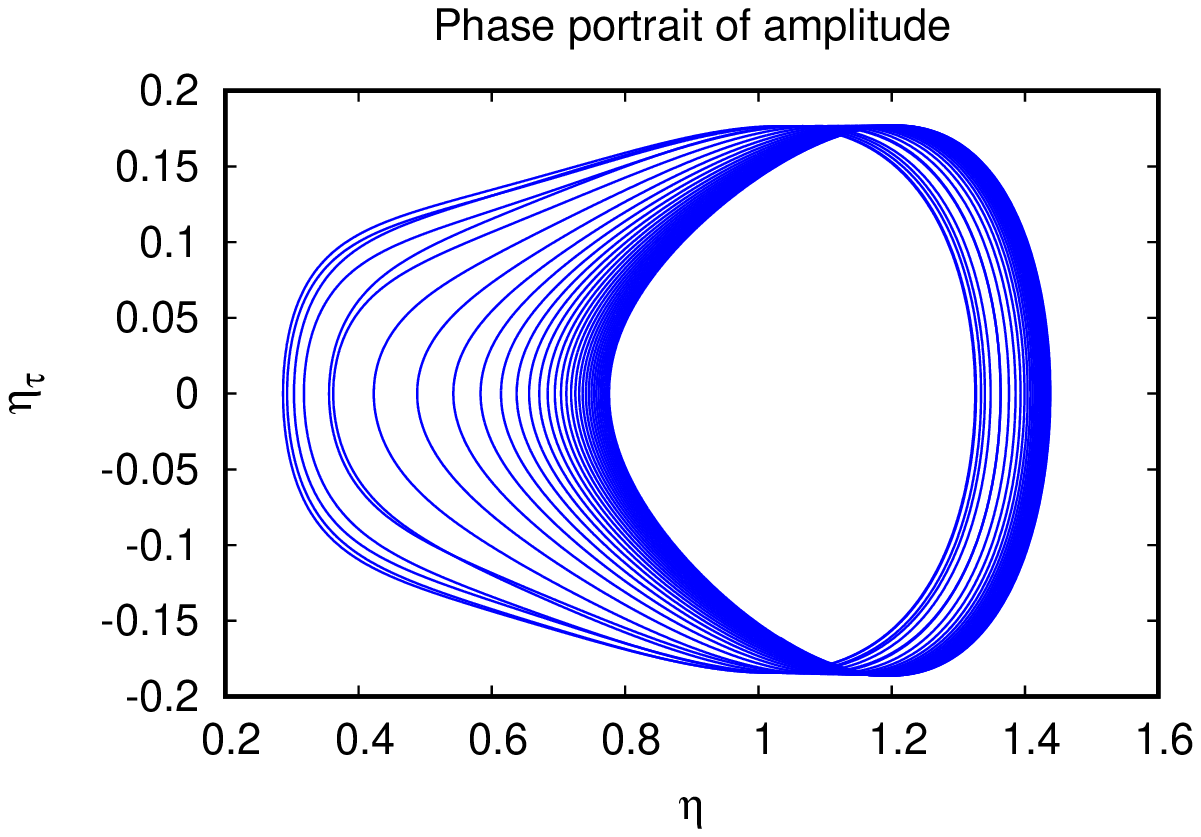}
\includegraphics[width=1.51in,height=1.3in]{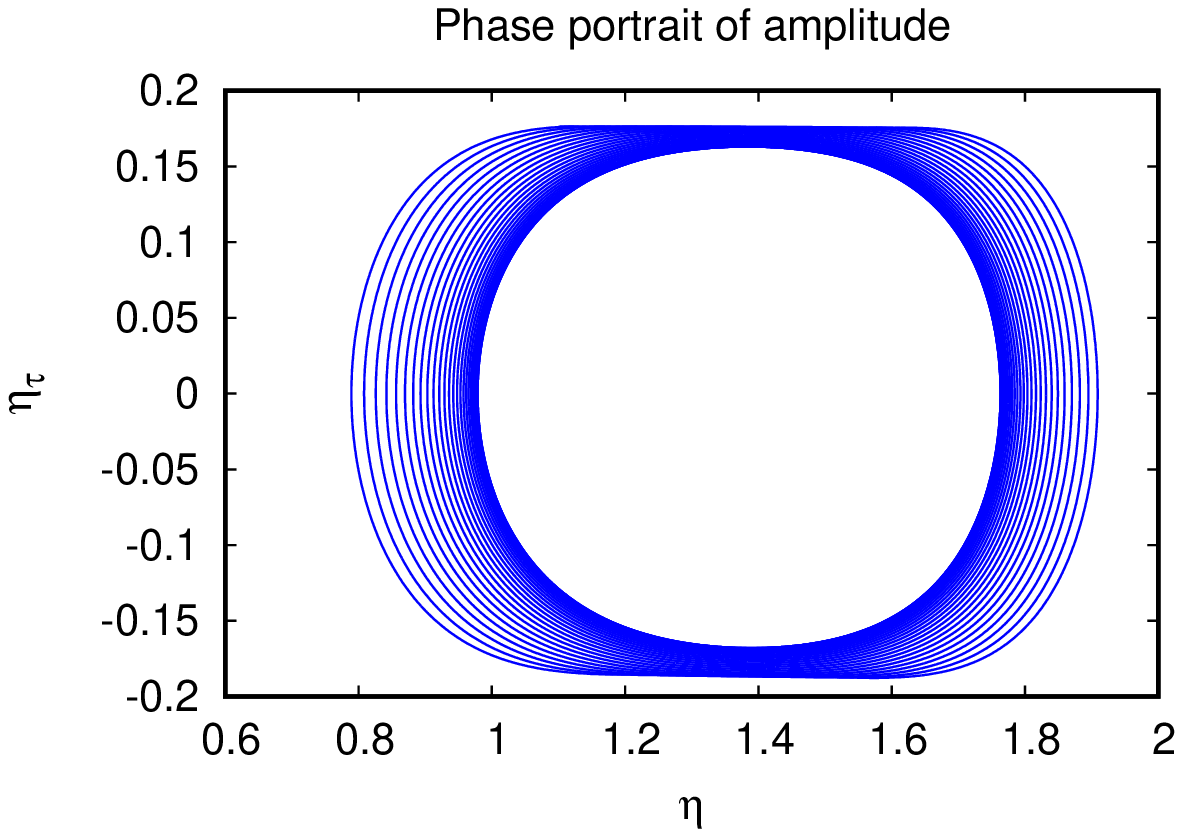}
\includegraphics[width=1.51in,height=1.3in]{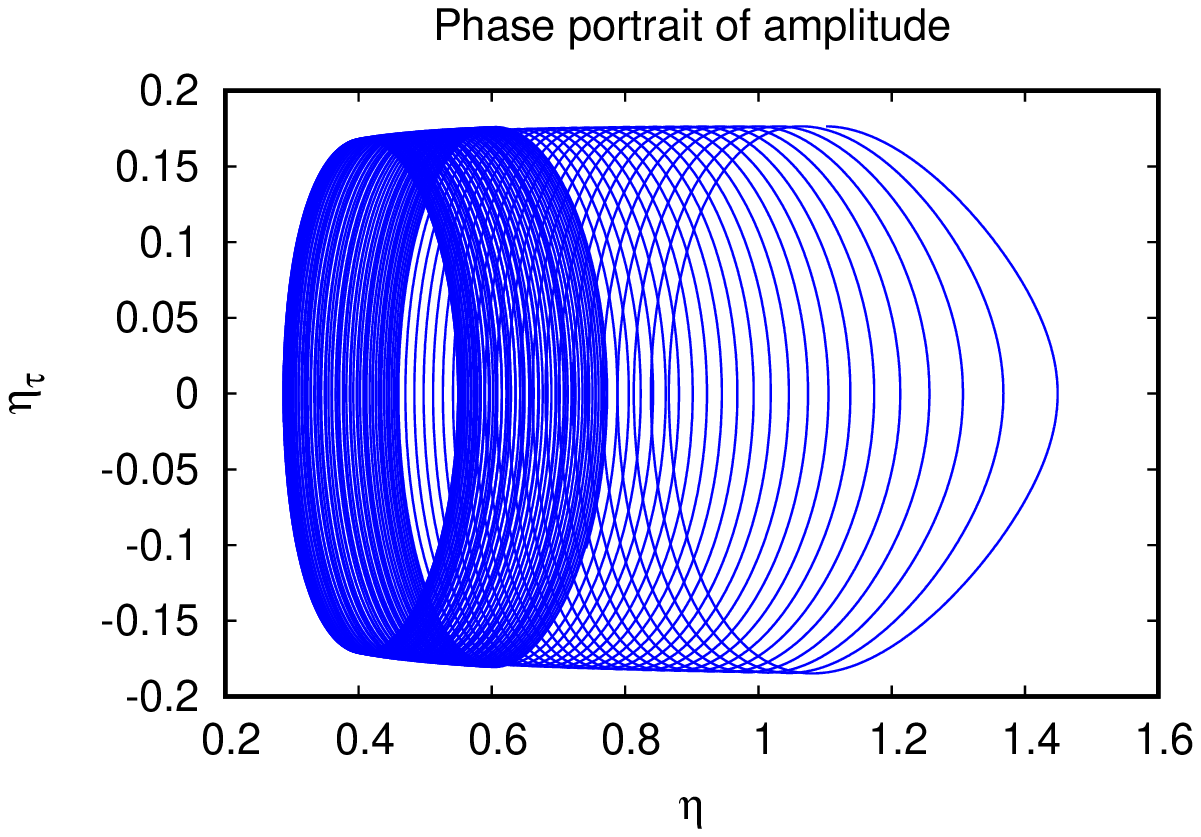}
\includegraphics[width=1.51in,height=1.3in]{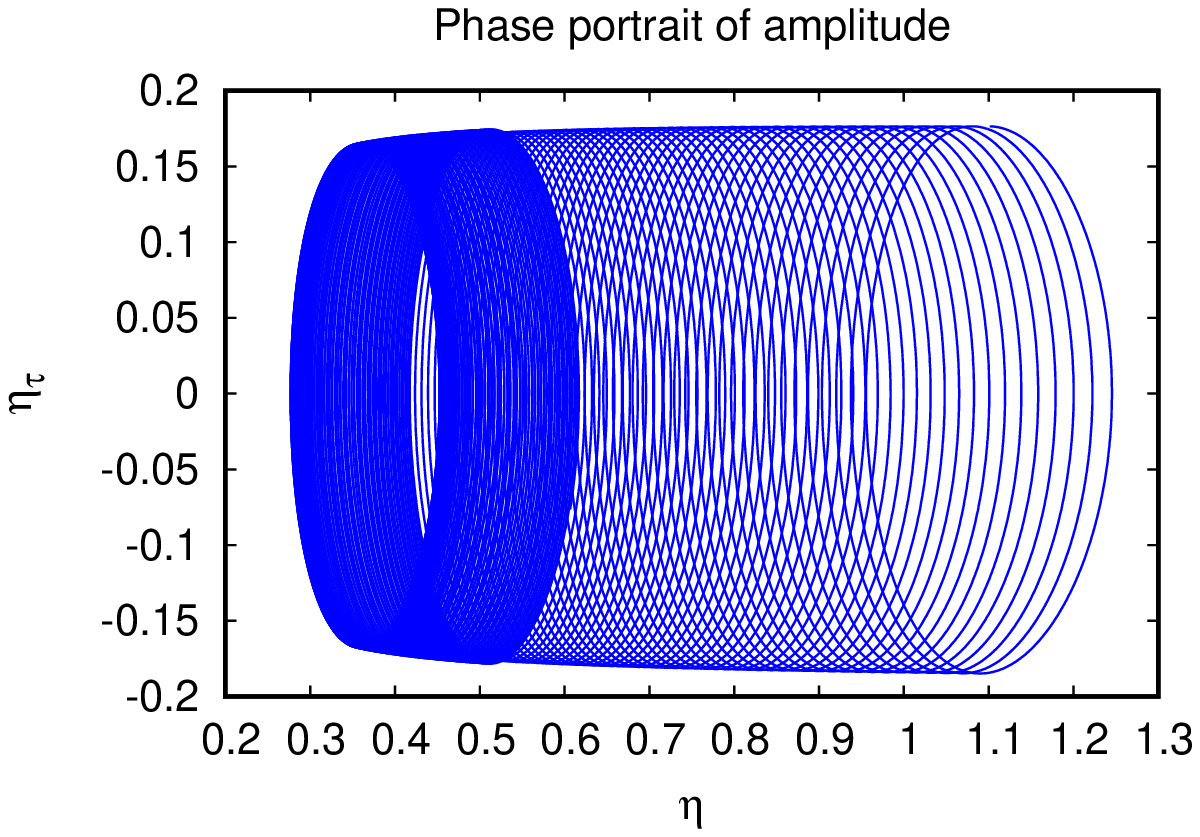}
\caption{\label{f7}(Color online) Phase portraits of the amplitude for $\alpha_1=0.001$. From top to bottom and from left to right: $\alpha_2=0.05$, $\alpha_2=0.5$, $\alpha_2=0.7$, $\alpha_2=1$.} 
\end{figure}

The apparent control of the Elliptic-soliton stability by the linear loss coefficient $\alpha_1$, has an important implication as concerns the dynamics of optical soliton frequency combs. Indeed, besides characterizing noise in the phase spectrum of the soliton and thus affecting its repetition frequency, the quality factor $Q$ (a dimensionless parameter describing how underdamped a resonator is) of the microresonator is inversely proportional to the microresonator loss coefficient. This means that the average lifetime of a comb structure in the soliton state will depend on the linear loss: when the linear loss of the cavity is increased, the soliton comb should fade away more or less shortly after its emission.  On the contrary, a decrease in the linear loss will enhance the signal intensity in the cavity. The later phenomenon is quite common in laser processings and is known as $Q$-switching. Note that at very high values of the linear loss $\alpha_1$, compared with the pump detuning frequency $\alpha_2$, the evolution of the soliton comb becomes chaotic as seen in fig.~\ref{f5}d.\\
The changes observed in fig. \ref{f6} on the time-series evolution of the Elliptic soliton amplitude, due to a variation of the pump detuning frequency $\alpha_2$, suggest a very rich dynamics of the system related to this parameter.
To this last point, fig.~\ref{f6} shows that as we increase $\alpha_2$ we move from regularly oscillating Elliptic-soliton structures to "rolling" patterns (panels (c) and (d) in fig.~\ref{f6}). This behaviour is more apparent by looking at the phase portrait of the amplitude, which we represented in fig.~\ref{f7} for a fixed value of $\alpha_1$. 

\begin{figure}[h]
\includegraphics[width=1.51in,height=1.3in]{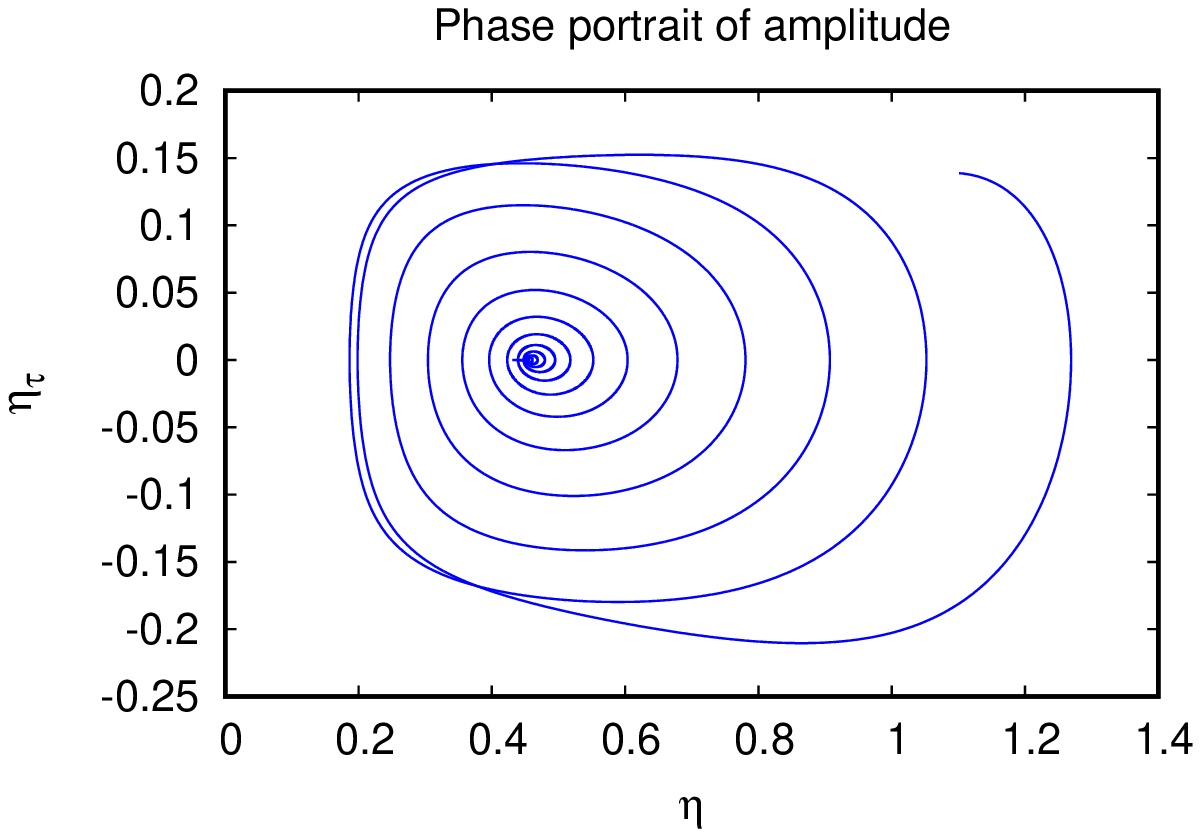}
\includegraphics[width=1.51in,height=1.3in]{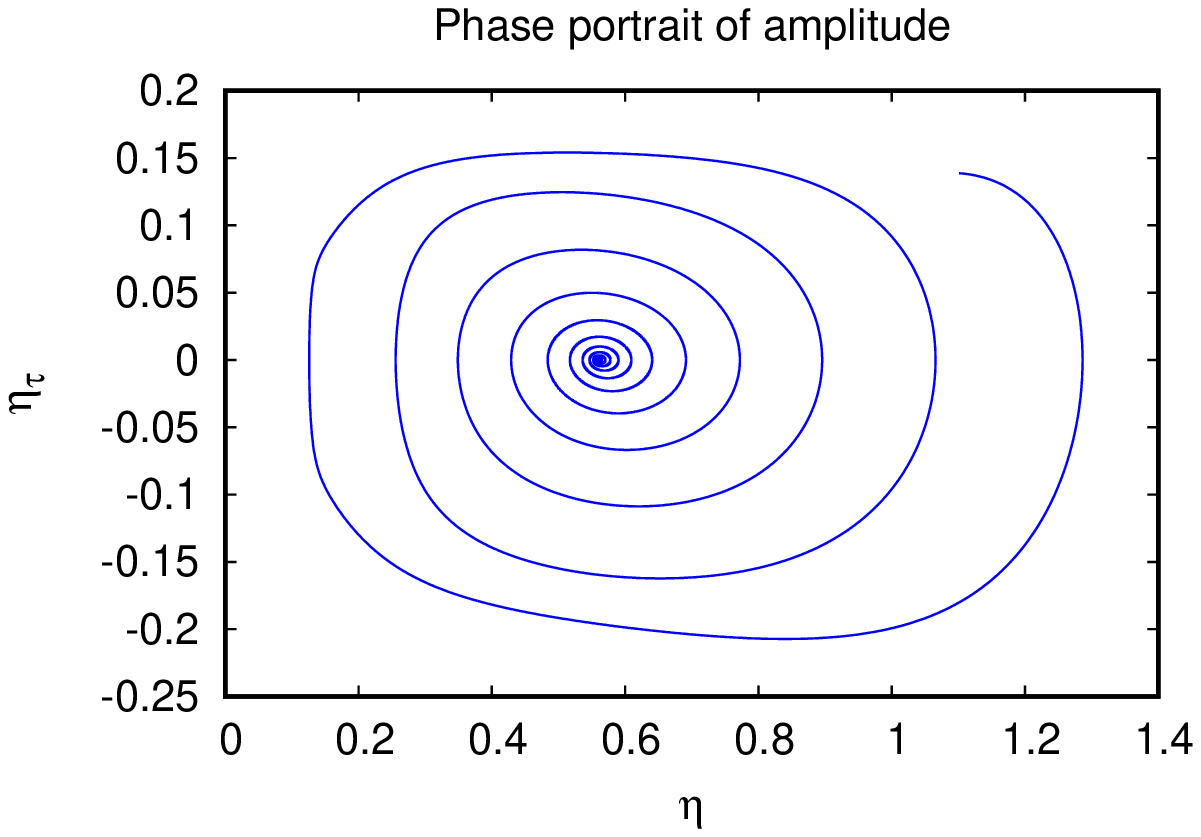}
\caption{\label{f8}
(Color online) Phase portraits of the amplitude for $\alpha_2=0.01$ and $\alpha_1=0.005$ (left), $\alpha_1=0.01$ (right).} 
\end{figure}

\begin{figure}[h]
\includegraphics[width=1.51in,height=1.3in]{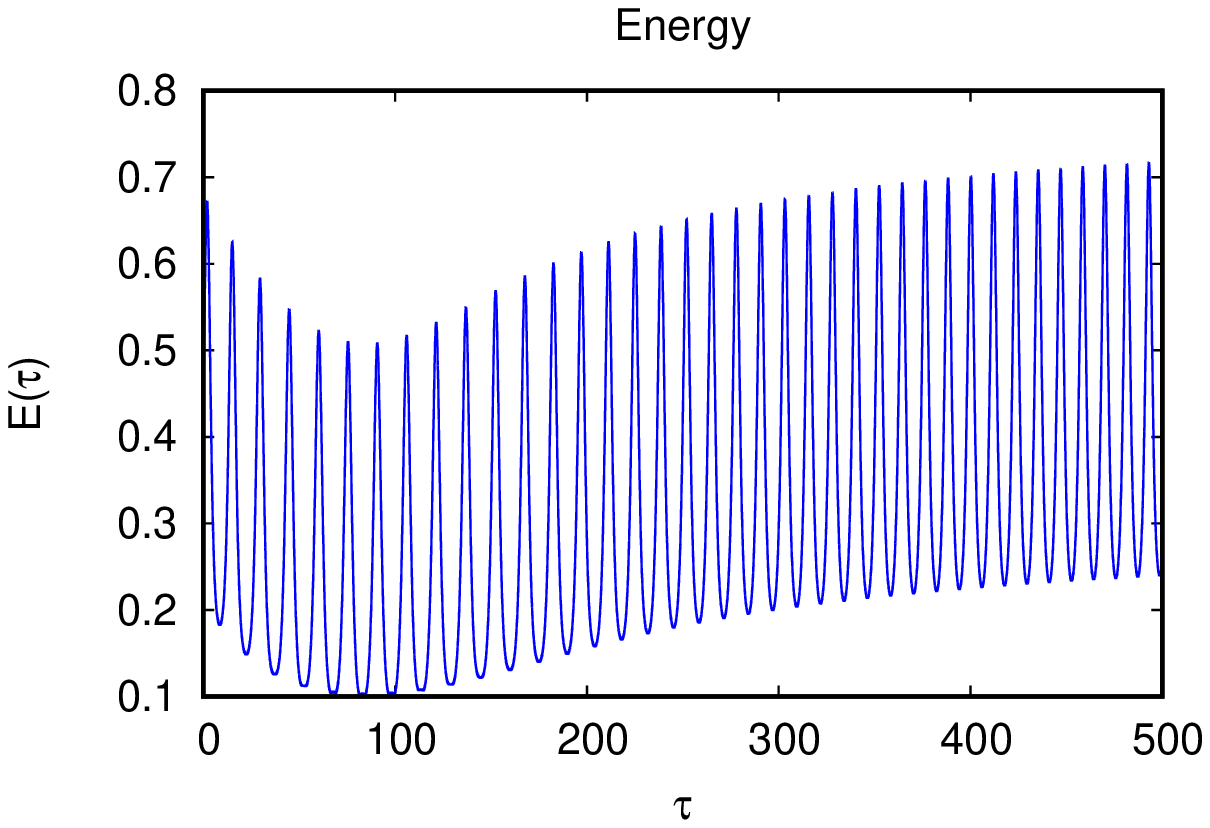}(a)\includegraphics[width=1.51in,height=1.3in]{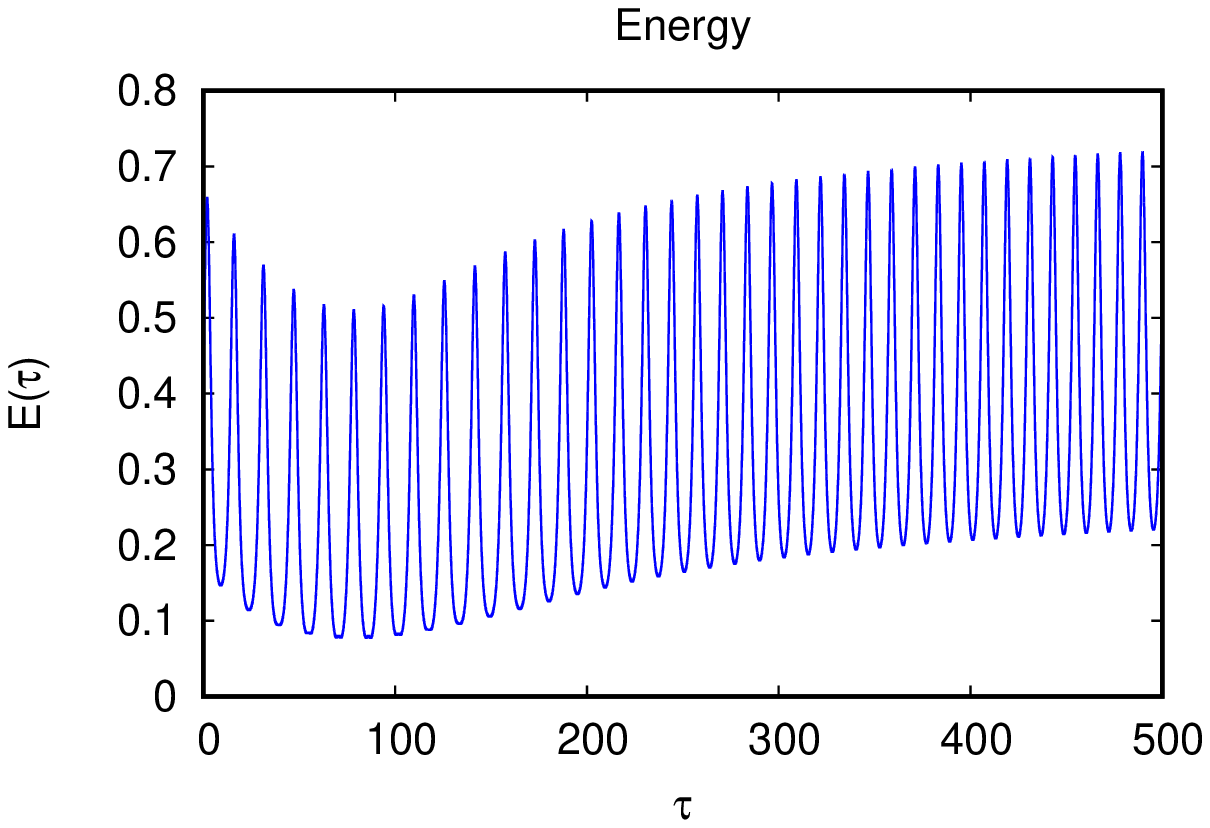}(b)
\includegraphics[width=1.51in,height=1.3in]{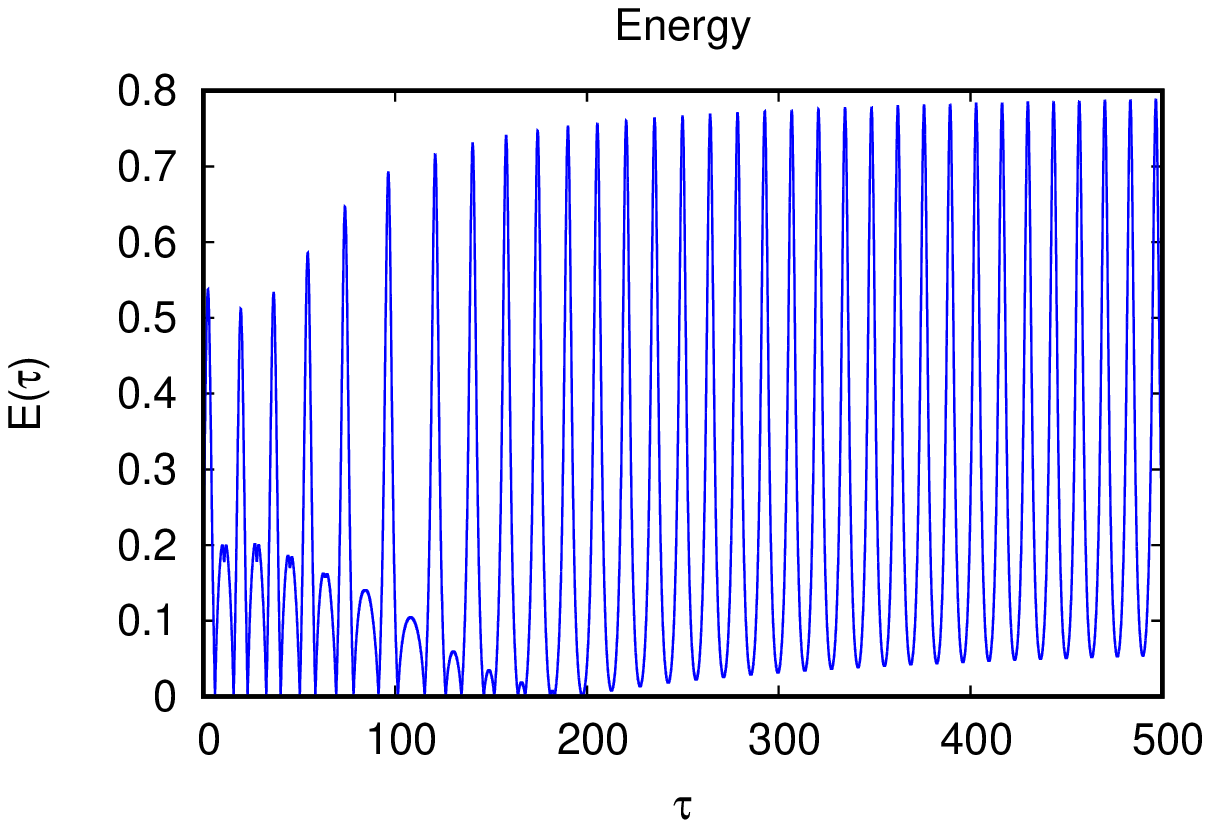}(c)\includegraphics[width=1.51in,height=1.3in]{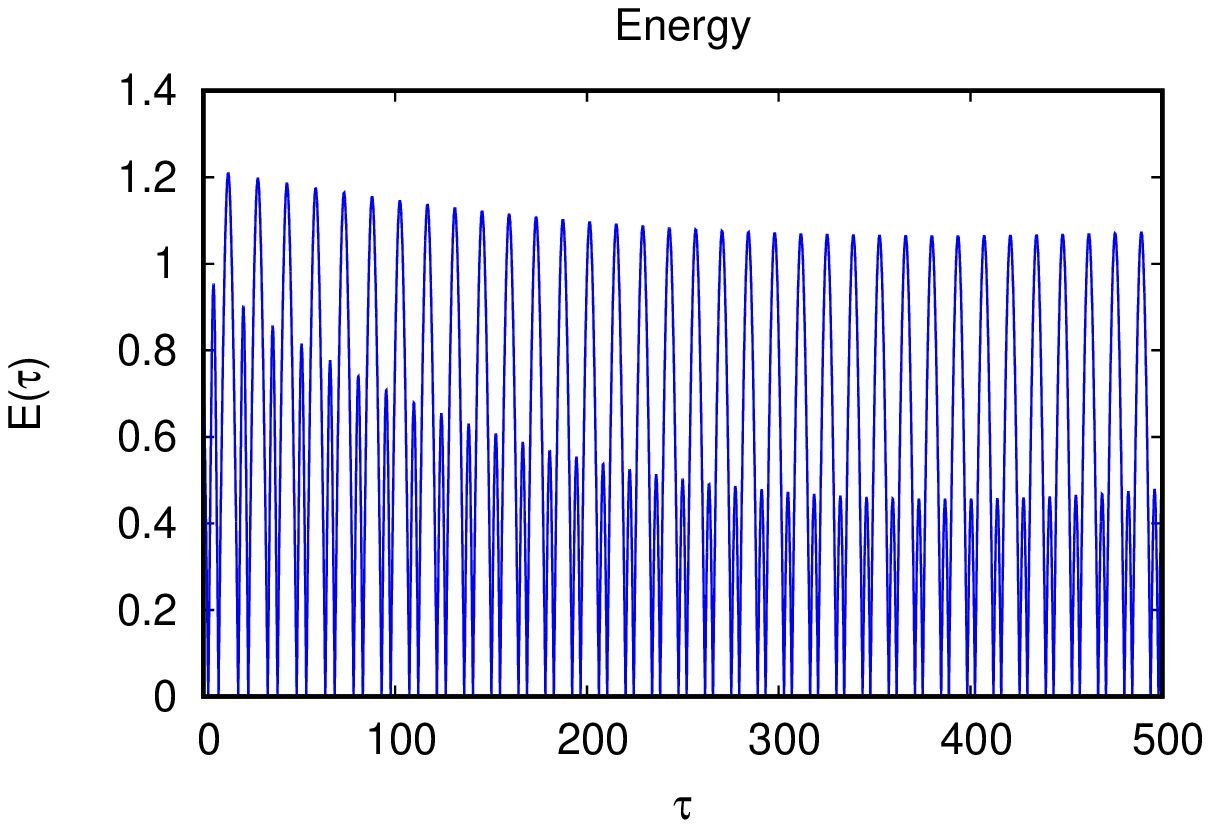}(d)
\caption{\label{f9}
(Color online) Time variation of the Elliptic-soliton energy. From graphs (a) to (f) the linear loss parameter is constant (i.e. $\alpha_1=0.001$), while the pump
detuning frequency is varied as: (a) $\alpha_2 =0.001 $, (b) $\alpha_2 =0.01$, (c) $\alpha_2 = 0.1$,
(d) $\alpha_2 = 0.5$.} 
\end{figure}

For values of $\alpha_2$ around $1$ and above, the three-dimensional representation of the Elliptic-soliton amplitude exhibits a spin-like behavior. Note that these patterns are in agreement with experiments~\cite{r3,r24,r25}, where it is observed that roll patterns are formed when the resonator is pumped above a certain threshold. \\
To provide a more global picture of the temporal evolution of the Elliptic-soliton amplitude in the LLE, we also plotted the phase portrait for varying loss coefficient$\alpha_1$ and fixed pump detuning frequency $\alpha_2$. 
Fig.~\ref{f8} are plots of these phase portraits. 
Essentially fig.~\ref{f8} suggests that when the coefficient of linear loss $\alpha_1$ is increased, the oscillatory feature of the Elliptic-soliton amplitude in the resonator cavity tends to die down, thus confirming the stabilizing role played by this parameter in the Elliptic-soliton dynamics. \\
In a final analysis we examine the time variation of the "topological" energy of the Elliptic soliton, considered as another relevant indicator of the soliton-comb stability as it propagates in the ring microresonator. In our analysis we consider numerical results only for variations of the pump detuning frequency, avoiding chaotic evolutions by fixing the linear loss coefficient in the regime of regular motion. \\
We derive the energy from the Lagrangian using the general relation: 
\begin{equation}
 \textit{E}=\left( \sum_{i} \dot{q}_i \frac{\partial }{\partial \dot{q}_i} - 1 \right) L,
 \label{e26}
\end{equation}
where $q_i$ is one of the collective-coordinate variables $\eta$, $q$, $\phi$, $\delta$. With the help of Eq.~(\ref{e22}), we find the energy:
\begin{eqnarray}
\textit{E}=\frac{4}{3}(2-\kappa^2) \eta^3 E(\kappa) -\eta [(2-\kappa^2)\eta^2 + \delta^2]E(\kappa)
\nonumber \\ + 2 Im\int^{K}_{-K}\varepsilon\psi^* d\vartheta.
\label{e27}
\end{eqnarray}
The time variation of the Elliptic-soliton energy is depicted in fig.~\ref{f9}, for some values of linear loss and the pump detuning frequency. 

The figures show that as one increases the pump detuning frequency, a second oscillatory curve shows up which overtakes gradually the Elliptic-soliton energy. This second energy may be attributed to the emergence of roll patterns in the system. When $\alpha_2>0.6$, the roll patterns completely wipe out solitonic structures in the system.

\section{\label{sec:6} Conclusion}

Optical frequency combs stem from outstanding progress achieved over the last two decades in precision control and stabilization of mode-locked ultrafast lasers. Soliton combs \cite{kart1,kart2,kart3} in particular have emerged as cavity solitons \cite{miro} consisting of a regular comb of sharp pulse lines, produced by mode-locked lasers from femtosecond optical frequency comb generators. These specific laser patterns have revolutionized optical frequency metrology and synthesis, they serve as basis for demonstrations of atomic clocks that utilize an optical frequency transition and have recently shown efficiencies in time-domain applications, including phase-sensitive extreme nonlinear optics and pulse manipulation as well as control \cite{r13,r26,r28,r29,r30}. \\
The aim of the present work was to propose an extensive study of the mechanism of generation and the dynamics of soliton combs in ring-shaped optical microresonators, within the framework of the Lugiato-Levefer equation. Being a specific form of perturbed cubic nonlinear Schr\"odinger equation, it is ready to assume that the solitonic feature of pulses composing the soliton comb arises from the NLSE while the perturbation, related to the cavity loss and the pump fields, are expected to determine the comb dynamics. Based on this consideration we first addressed the issue of the generation of soliton-crystal structures that have recently been observed in monolithic Kerr optical frequency comb microresonators. We established that these structures, which have previously been represented as a periodic train of spatially entangled pulses, are equivalent to the Elliptic-soliton solution to the homogeneous NLSE governing equally their pulse components. We analyzed the stability of the Elliptic soliton and obtained a rich boundstate spectrum consisting of unstable as well as stable long-term internal oscillations existing in the Elliptic-soliton backgound. A collective-coordinate approach to the Lugiato-Lefever equation was developed, and numerical simulations were carried out to point out the effects of loss and pump detuning on the Elliptic-soliton profiles. In particular we found that while the time evolution of the Elliptic-soliton amplitude is dominantly oscillatory, a variation of the two perturbation parameters gives rise to a quite rich dynamics including rolling patterns and chaotic evolutions. \\
The values of $\alpha_1$ and $\alpha_2$ that were considered in the numerical analysis of sec.~\ref{sec:5}, actually underscore only a very little aspect of the extremely rich dynamics of the system. Nevertheless the few values considered provide a valuable insight onto the system dynamics, as reflected in time series of the Elliptic soliton amplitude and the associate phase portraits, as well as the time variation of the Elliptic-soliton energy. 

\acknowledgments
The authors thank the Alexander von Humboldt (AvH) foundation for logistic supports.

\end{document}